\documentstyle[11pt,aaspp4,tighten]{article}

\lefthead{Jones, Ryu, \& Tregillis}
\righthead{MHD Cosmic Bullets}

\begin{document}

\font\twelvei = cmmi10 scaled\magstep1 
       \font\teni = cmmi10 \font\seveni = cmmi7
\font\mbf = cmmib10 scaled\magstep1
       \font\mbfs = cmmib10 \font\mbfss = cmmib10 scaled 833
\font\msybf = cmbsy10 scaled\magstep1
       \font\msybfs = cmbsy10 \font\msybfss = cmbsy10 scaled 833
\textfont1 = \twelvei
       \scriptfont1 = \twelvei \scriptscriptfont1 = \teni
       \def\mit{\fam1 }
\textfont9 = \mbf
       \scriptfont9 = \mbfs \scriptscriptfont9 = \mbfss
       \def\bmit{\fam9 }
\textfont10 = \msybf
       \scriptfont10 = \msybfs \scriptscriptfont10 = \msybfss
       \def\bmsy{\fam10 }

\def\etal{{\it et al.~}}
\def\eg{{\it e.g.}}
\def\ie{{\it i.e.}}
\def\lsim{\raise0.3ex\hbox{$<$}\kern-0.75em{\lower0.65ex\hbox{$\sim$}}} 
\def\gsim{\raise0.3ex\hbox{$>$}\kern-0.75em{\lower0.65ex\hbox{$\sim$}}} 

\title{The Magnetohydrodynamics of Supersonic Gas Clouds:\\
    MHD Cosmic Bullets and Wind-Swept Clumps\altaffilmark{7}}

\author{T. W. Jones\altaffilmark{1,4},
        Dongsu Ryu\altaffilmark{2,5},
        and I. L. Tregillis\altaffilmark{1,3,6}}

\altaffiltext{1}{Department of Astronomy, University of Minnesota,
Minneapolis, MN 55455}
\altaffiltext{2}
{Department of Astronomy \& Space Science, Chungnam National University,
Daejeon 305-764, Korea}
\altaffiltext{3}{present address: Department of Applied Physics,
Cornell University, Ithaca, NY 14853}
\altaffiltext{4}{e-mail: twj@astro.spa.umn.edu}
\altaffiltext{5}{e-mail: ryu@sirius.chungnam.ac.kr}
\altaffiltext{6}{e-mail: tregilli@msi.umn.edu}
\altaffiltext{7}{Submitted to the Astrophysical Journal}

\begin{abstract}

We report an extensive set of two-dimensional MHD simulations exploring
the role and evolution of magnetic fields in the dynamics of supersonic
plasma clumps.
We examine the influence of both ambient field strength and
orientation on the problem. Of those two characteristics, field
orientation is far more important in the cases we have considered
with $\beta_0 = p_g/p_b \ge 1$. That is due to the geometry-sensitivity of field
stretching/amplification from large-scale shearing motions around the
bullet.
When the ambient magnetic field is transverse to the bullet motion,
even a very modest field, well below equipartition strength, can be 
amplified by field line stretching around the bullet within a couple 
of bullet crushing times so that Maxwell stresses
become comparable to the ram pressure associated with the bullet motion.
The possibility is discussed that those situations might lead to 
large, induced electric potentials capable of accelerating charged particles.
When the ambient field is aligned to the bullet motion, on the other hand, 
reconnection-prone topologies develop that shorten the stretched field
and release much of the excess energy it contains.
In this geometry, the Maxwell stresses on the bullet never approach the
ram pressure level.
In both cases, however, the presence of a field with even moderate initial strength
acts to help the flow realign itself around the bullet into a smoother,
more laminar form. That reduces bullet fragmentation tendencies
caused by destructive instabilities. Eddies seem less effective at
field amplification than flows around the bullet, because fields
within eddies tend to be expelled to the eddy perimeters. Similar effects
cause the magnetic field within the bullet itself to be reduced below its initial
value over time.

For oblique fields, we expect that the transverse
field cases modeled here are more generally relevant. What counts is whether field
lines threading the face of the bullet are swept around it
in a fashion that folds them (leading to reconnection) or that keeps them
unidirectional one each side of the bullet. In the second instance, behaviors should resemble those of the
transverse field cases. We estimate that this, quasi-transverse, behavior
is appropriate whenever the angle, $\theta$, between the motion and
the field satisfies $tan{\theta} \gsim 1/M$, where $M$ is the bullet
Mach number.

From these simulations, we find support in either field geometry
for the conclusions reached
in previous studies that nonthermal radio emission
associated with supersonic clumps is likely to be controlled largely 
by the generation of strong  magnetic fields around the perimeters
of the clumps, rather than 
local particle acceleration and field compression within the bow shock. 
In addition, since the magnetic pressure on the nose of the bullet likely
becomes comparable to the ram pressure and hence the total pressure
behind the bow shock, the gas pressure there
could be substantially lower than that
in a gasdynamical bullet. That means, as well, that the temperature
in the region on the nose of the bullet would be lower than that
predicted in the gasdynamical case. That detail could alter
expectations of the thermal emission, including X-rays and UV-IR
lines.

\end{abstract}


\clearpage

\section{Introduction}

In recent years observations have revealed
the presence of dense, high-velocity
gas clumps within and sometimes outside young supernova remnants and also
associated
with the outflows from 
young stars (\eg, \cite{braun87}; \cite{allen93};
\cite{asch95}; \cite{strom95}).
These ``cosmic bullets''
or, in some scenarios, ``wind-swept clumps'' can shine prominently
in either thermal emission (\eg, \cite{vanb71}; \cite{schwar75};
\cite{allen93}; \cite{strom95}) or 
nonthermal emission (\eg, \cite{braun87}; \cite{ander94}). Sometimes in
supernova remnants they show elemental compositions indicating 
formation exclusively from material ejected out of the depths of
the progenitor supernova. The exact origins of the bullets
are still unclear, but suggestions range from explosive ejection
(\cite{loeb}; \cite{norsilk79}) to wind-swept,
pre-existing clumps (\cite{schian95}),
and dynamical instabilities in winds (\cite{ston95}).
Whatever the manner of creation, cosmic bullets seem capable of surviving
for remarkable lengths of time, thus transporting
energy and mass over large distances, and have been suggested as sites
for acceleration of high energy particles (\eg, \cite{colbick85};
\cite{braun87}; \cite{jonkant}). Yet, simple
arguments and computer simulations suggest that
gasdynamical instabilities may be highly disruptive
of these bullets (\eg, \cite{jonkant}; \cite{schian95}), limiting their
existence to only a few times the interval required
for the applied ram pressure
to crush them.
That would limit their propagation distance to only a few times the length
$\sqrt{\chi} R$,
where $\chi$ is the density contrast with the environment and $R$ is the
bullet radius. So, only the densest clumps could survive to propagate
more than a few times their own size scale.

However, in many cases one expects that both the clouds
and the surrounding material are mostly ionized plasmas,
so that embedded magnetic fields ought to
be largely frozen in and should play a role.
In a recent paper we emphasized the possibility, based on numerical
simulations involving a passive magnetic field, that bullets may be effective
in amplifying weak ambient magnetic fields as a side effect of the destructive
interactions with their environments (\cite{jonkant}). It was
emphasized there that field amplification
could be more important in generation of associated nonthermal
synchrotron emission, for example, than local particle acceleration
in the shocks formed by the bullets (\cite{jonkan93};
\cite{ander94}; \cite{jonkant}). At the
same time amplified magnetic fields could act to stabilize the
Kelvin-Helmholtz and Rayleigh-Taylor instabilities, thereby limiting
bullet fragmentation. That role might help account for their ability to
survive, or at least for significant pieces to penetrate much farther than
gasdynamical calculations suggest they might.

\cite{maclow94} carried out two-dimensional MHD simulations of
a dense plasma cloud overrun by a plane shock, demonstrating
in that related case how a magnetic field might be amplified and also how
it may prevent the complete destruction of the cloud in the postshock
flow. Although
shocked clouds and supersonic bullets are qualitatively similar objects
in some ways, there are also
significant differences between them, as well.
In particular, as long as the impinging shock is not radiative,
the material surrounding the shocked cloud is mostly moving sub-sonically
with respect to the cloud, because 
the cloud is overrun by a hot, postshock flow. The bullet, of course, is
(by definition) interacting with a supersonic flow, at least in the
beginning. This difference influences especially the flow to the rear and
on the sides of the cloud and could influence instabilities and
magnetic field behavior (see \cite{jonkan93,jonkant}). Furthermore,
\cite{maclow94} considered primarily shocked clouds with
a magnetic field aligned along the symmetry axis of the flow, except for
one simulation involving a transverse field
that was not reported in much detail. 
That aligned field orientation was mandated 
by their use of cylindrical symmetry for all but the one simulation.
The other simple field geometry possible with cylindrical symmetry; namely, a 
toroidal one, is not a very natural
one for this problem. On the other hand, the geometry of the
magnetic field ought to be very important to its evolution in and
around the dense cloud and also to its dynamical influence on the cloud.
We may also expect the influence of the magnetic field to depend on its
strength. These issues in the bullet context have not been addressed 
before in the literature.

In this study we present results from two-dimensional,
full MHD simulations of the evolution
of dense supersonic plasma bullets with magnetic field geometries
both aligned with the bullet motion and also transverse to it.
In order to have freedom in the field geometry, the simulations have
been done in Cartesian coordinates rather than
in cylindrical coordinates.
Also we have simulated flows using a range of magnetic field strengths and
using a range of numerical resolutions. 
One can envision the bullets either as ejecta or as condensations
being swept by a supersonic wind.
Our objectives are to understand better how the field in and around the
bullet is influenced by the bullet and also how the magnetic field
acts on the bullet dynamically. \S 2 outlines the important
dynamical issues, while \S 3 describes the numerical
methods we employ. In \S 4 we describe our results,
while a brief summary of the principal conclusions is presented
in \S 5.

\section{The Dynamical Problem}

The basic issues associated with the interaction between a gas bullet and
its surroundings are well-known, so readers are referred to the extant 
literature for detailed discussions (\eg, \cite{jonkant}; \cite{schian95} and
references therein) or
for comparisons with shocked clouds (\eg, \cite{jonkan93};
\cite{klein94}; \cite{maclow94}; \cite{xustone}).
In this section we present only a brief
outline of the issues to facilitate later evaluation of
the particular new issues raised in this study.

\subsection{Problem Definition}

As in previous studies we consider the bullet to be initially in
pressure equilibrium with the ambient medium and the initial
magnetic field to be uniform through both the bullet and the ambient
medium. 
The plasma is assumed throughout to be highly conducting,
so that magnetic field is frozen
in. At $t = 0$ the bullet is set into motion at speed $u_{b0}$ with
respect to the ambient medium.
If such complications as radiative cooling, thermal conduction, and
gravity can be neglected, the bullet dynamics is determined by a minimum
of the following six parameters and the bullet shape:
the sonic Mach number of the bullet through the ambient
medium, $M = u_{b0}/c_{sa}$; the gas adiabatic index,
$\gamma$; the density contrast between the bullet and the ambient
background, $\chi = \rho_b/\rho_a$; as well as the magnetic field
strength, $B_0$,  and orientation (generally, two angles).
The initial strength of the
magnetic field can be expressed either in terms of the Alfv\'enic
Mach number $M_A = u_{b0}/a$, where $a$ is the Alfv\'en speed, or in
terms of $\beta_0 = p_g/p_b = (2/\gamma) (M_A/M)^2$.
The field orientation can be specified by the angle, $\theta$, between the field,
${\bmit B}_0$, and the bullet velocity, ${\bmit u}_{b0}$, as well as
between ${\bmit B}_0$ and any distinct symmetry axis
or plane for the bullet.

Fortunately, as previous shocked-cloud studies emphasized
(\cite{klein94}; \cite{maclow94}),
in the limit $M>>1$, the sonic Mach number can be scaled out of the problem.
\cite{xustone} found in three-dimensional gasdynamical
simulations of shocked clouds
that the initial morphology of the cloud does make a quantitative difference
in its evolution, but that qualitative features
are not very geometry-sensitive.
Thus, we will consider only a single initial Mach number and a single
bullet morphology. 
Because three-dimensional MHD simulations of
sufficient numerical resolution are
still too expensive to follow magnetic fields
adequately in this problem, we have
used two-dimensional symmetry.
We are concerned, on the other hand,
with influences of different magnetic field geometries with respect
to the bullet and its motion, so we have chosen a Cartesian coordinate
system, which enables us to set up a uniform magnetic field with 
arbitrary orientation.
In this geometry the simplest initial cloud morphology is a circular
cylinder, with its axis out of the computational plane and thus
orthogonal to the motion of the bullet.  Although we have actually
carried out test simulations in which the magnetic field has a component
out of the plane (and, thus, aligned with the bullet axis), we present
here only cases for which the magnetic field is entirely in the
computational plane and either aligned to the direction of the
bullet motion or transverse to it. We have found no qualitatively different
behaviors for the fields with a component out of the plane.

The characteristic evolutionary timescale of a bullet is
the so-called ``bullet crushing time'',
\begin {equation}
t_{bc} = {{2 R \sqrt{\chi}}\over{u_{b0}}},
\label{tcrush}
\end{equation}
where $R$ is the initial bullet radius.  It allows one to subsume
the values of the initial density ratio, $\chi$, and speed, $u_{b0}$, into
one parameter.
It is traditional in studies such as this to consider a single, 
convenient value for each of
these, and to discuss the evolution as though it depended only on
$t/t_{bc}$ and not on $\chi$, for example. We will follow this
procedure, but also comment on its limits. In addition 
we set $\gamma = 5/3$. We are
left, then, with a two-dimensional parameter space to explore; namely,
we need to define the strength and the orientation of the magnetic field
relative to the bullet motion in the plane.

\subsection{Issues}

Once set in motion, the bullet history begins through
formation of a shock pair; namely,
a strong bow shock that surrounds the bullet and an equally strong,
internal or ``bullet shock'',
that propagates through the bullet from nose to tail
compressing it along the direction
of motion. The time, $t_{bc}$, defined in Eq.~\ref{tcrush}
measures the interval required for the ram-pressure-driven bullet
shock to cross the bullet
diameter. Note that our definition for $t_{bc}$ contains an
additional factor of two compared to that normally
used in discussions of shocked clouds.
Before $t_{bc}$, the bow shock is established and flows along
the bullet boundary may lead to some stripping
due to small-scale Kelvin-Helmholtz (K-H) instabilities.
On a timescale comparable to or exceeding $t_{bc}$, both K-H
and Rayleigh-Taylor (R-T) instabilities with scales comparable to the
size of the bullet can begin to develop.
Gasdynamical calculations mentioned earlier show
R-T instability-induced bubbles with low density, ambient plasma 
penetrating through the bullet body, beginning a fragmentation process.
K-H instabilities along boundaries allow small pieces of the bullet
to be stripped, so eventually one expects the gas bullet to be broken into
a ``mist'' and then dispersed. Estimates for this ``dispersal'' time are
consistent for both bullet and shocked cloud simulations. Although
the dispersal timescale is difficult to define uniquely,
all gasdynamical simulations show
significant signs of bullet disruption by about $t \sim 2-3 t_{bc}$
(using our definition for the time unit).
Before the bullet is fully disrupted, ram pressure decelerates it
substantially (or if it is seen as a wind-swept clump, accelerates
it in the direction of the wind).
That deceleration, of course, is the origin of the R-T
instability. The deceleration is considerably enhanced by lateral expansion of
the bullet coming from the penetration of the bubbles induced by R-T
instabilities.

The simulations presented by \cite{jonkant} and \cite{maclow94} showed
that weak magnetic fields in the ambient medium can be stretched and
thus strengthened through sheared gas motions around clouds.
Compression can also increase field strength, but that process is
generally much less significant in complex flows than stretching.
\cite{maclow94} concluded from their shocked cloud simulations
that the fields can sometimes be amplified until $p_b \sim \rho_a u^2_b$.
Field enhancement through stretching takes place in three situations.
\cite{maclow94}
emphasized the importance in their axial geometry of field
lines aligned along the symmetry axis being
stretched along the symmetry-axis, behind their shocked cloud, while a Mach disk
carries away (conducting) material from the cloud as it
catches up to the originally incident shock in recession.
\cite{jonkant} emphasized in their transverse field geometry the
importance of field lines being stretched around the cloud perimeter
as they envelope the cloud.
In addition, for either field geometry, vortices that form, especially
in the immediate wake of the cloud, could stretch entrained field lines 
and, thus, amplify them. 
Field amplification within vortices is complex, however, because it
involves field lines that are wrapped around until they are adjacent to
fields of opposite direction. That is unstable to reconnection, so
that within the vortices field line reconnection 
will limit the field growth and effectively expel magnetic flux (\eg,
\cite{weiss66}; \cite{gallweiss81}) to the vortex perimeter. For numerical
simulations of ideal MHD, the resistivity that allows reconnection to
take place comes from numerical truncation and the associated
dissipation, so it is locally resolution dependent.
That means one needs to be cautious in understanding the convergence
properties of fields. We emphasize, however, that with the fully conservative
methods we employ (including mass, momentum, total energy and magnetic
flux), the reconnective process is a necessary consequence, not an added feature.

\cite{maclow94} found from their shocked cloud simulations that 
inclusion of a dynamical magnetic field did significantly enhance
the durability of their clouds, and concluded that this was
primarily due to the influence of the magnetic field in inhibiting
the vortices that form along the slip surfaces on the sides and rear
of the cloud. Although \cite{maclow94} defined
their initial field strength so that $\beta_0 = 1$ for most of the runs
they did, this was based on the {\it preshock} ambient plasma. Behind
the incident (Mach 10) shock, where the interactions are taking place,
$\beta_0 \propto M^2 >>1$.
Thus, the field interacting with the clouds is initially relatively
weak.
A similar consideration applies to the MHD bullet problem, especially in the cases
with the field aligned to the bullet motion, since 
the value of $\beta$ behind the bow shock of the bullet is also
increased by a factor $\sim M^2$.

\section{Numerical Method \& Setup}

Our simulations have been carried out using a code for compressible,
ideal MHD based on a conservative, explicit TVD method as described
in \cite{ryujon95} and \cite{ryujonf}.
The scheme, utilizing an approximate MHD Riemann solver, is second 
order accurate in both space and time and cleanly captures all the
various families of MHD discontinuities. We have used the multidimensional,
Cartesian version of the code that was previously applied successfully
to the study of the nonlinear MHD K-H instability in modestly
compressible conditions (\cite{franetal}),
a problem that demonstrates the code's ability to follow accurately
smooth but complex MHD flows.
The code maintains ${\bmsy\nabla}\cdot{\bmit B}=0$ to machine accuracy
by employing a simple, corrective transformation at each time step
(\cite{brackbar})
utilizing an exact Fourier transform relationship as a key feature
to maintain both accuracy and speed in this step (\cite{ryujonf}).
The simulations are two and a half $(2+1/2)$ dimensional;
\ie, vector fields include three components,
but all fields are independent of the $z$-coordinate. The results
presented all involve flows with $B_z = u_z = 0$, however.

In each presented computation the bullet is
placed with its diameter along the x-axis, and reflection
symmetry assumed across that axis, so that only the upper half of the
flow structure has been computed. We have carried out several comparison test
simulations, however, using the full flow to demonstrate an exact 
correspondence with the more economical computations presented. 
The top and right boundaries of the grid are open, while the left
boundary is ``flow-in'' to maintain the upstream wind conditions exactly.

The bullet initially has a circular cross section
with a radius $R = {50}/{51.2} = 0.97656$. This, slightly noninteger value resulted from our
definitions of the bullet radius in zones (\eg 50) that did not quite match 
onto the grid sizes that we used (\eg 512).
Inside this radius
the bullet density is $\rho_b = \chi\rho_a$ with $\rho_a = 1$.
A thin boundary layer
with a hyperbolic tangent profile has been applied to the bullet,
using a characteristic width of two zones.
Since the grid is Cartesian, the
bullet has a cylindrical form and its axis is aligned with the $z$-direction.
The computational domain
is $x = [0,10]$, $y = [0,5]$. At the start of each computation the
bullet center is $(x_c, y_c) = (1.5, 0)$.  Each simulation
begins with the bullet at rest while an ambient medium flows from
left to right at speed $u_{b0} = 10 = M$, so that the sound speed in the
ambient medium, $c_{sa} = \sqrt{\gamma p_0/\rho_a}$, is unity. 
Interaction with this ``wind'' accelerates the
bullet towards the positive $x$-direction.
In a fixed frame this would cause the bullet to be expelled from the
grid on a timescale of only $\sim 2 t_{bc}$.
In order to enable the 
simulations to follow the bullet evolution as long as possible, we have
incorporated a feature that adjusts the reference frame speed
at each time step
so that the intersection of the bullet bow shock with the $x$-axis
is approximately stationary on the grid.
The bullet still moves slowly towards the right grid boundary as
the Mach number of the bow shock decreases and the stand-off distance
between the shock and the bullet nose increases.
But this procedure has enabled us to follow bullet
evolution for at least $4~t_{bc}$ and more typically $8~t_{bc}$.

We have used a range of numerical resolutions. Following \cite{klein94},
we have characterized them in terms of the number of zones across one bullet
radius; \eg, $R_{50}$, refers to a simulation with 50 zones across
the bullet radius. Since the grid is uniform the total number of zones
would be $512\times256$ for an $R_{50}$ run and
$1024\times512$ for an $R_{100}$ run, for examples.

To understand better the evolution of the bullet,
we have included a passive, Lagrangian,
conserved ``tracer'', $f$, that can be termed the ``bullet fraction''
or, in the language of \cite{xustone}, the ``color'' of the fluid. This
is set equal to unity for fluid initially inside the bullet $(f = 1)$
and zero $(f = 0)$ everywhere else.
The bullet fraction is followed with a TVD advection routine identical
to that utilized for mass advection in the code.
This enables us to compute several useful quantities, such as the
bullet mass inside the grid,
\begin{equation}
M_{bull} = \int_V f \rho dV,
\label{mbull}
\end{equation}
where $dV = dxdy$ and $V$ represents the volume of
the entire calculation domain. 
Other related useful quantities include
the mass-weighted speed of the bullet,
\begin{equation}
\bar u_b = {1\over M_{bull}}\int_V u_x f \rho dV,
\label{uxav}
\end{equation}
and the mean height of the bullet in the $y$-direction,
\begin{equation}
\bar h_{bul} = {1\over M_{bull}}\int_V y f \rho dV.
\label{yav}
\end{equation}
We will present $\bar u_b$ as measured in the initial rest frame
of the bullet.
At the start $\bar h_{bul} = 4 R/3\pi = 0.414$.
Another useful integrated quantity is the magnetic
energy enhancement inside the grid. That is conveniently
normalized in terms of the magnetic energy initially inside the bullet,
$E_{Bob} = (3/5) (\pi R^2/2 \beta_0)$, where we set $c_{sa} = \rho_a = 1$,
and $\gamma = {5}/{3}$.
Then using the initial total magnetic energy inside the computational domain,
$E_{Bo} = (3/5) (5\times 10/\beta_0)$, we have
\begin{equation}
\Delta E_{mag} = {1\over E_{Bob}}{\left [{\int_V {1\over 2}(B_x^2~+~B_y^2) dV
- E_{Bo}} \right ]},
\label{emag}
\end{equation}
where $p_b = \case{1}{2} B^2$ in our units.

\section{Results}

As mentioned earlier we have carried out simulations with two initial
magnetic field geometries; namely a field transverse to the motion of the
bullet (henceforth identified as ``T'' models) and parallel
or aligned with the bullet motion (henceforth identified
as ``A'' models). Table~\ref{tbl-1} characterizes 
six MHD simulation pairs with a range of field
strengths for each geometry.
Models T1--T5 and A1--A5 include 50 zones across a radius
(thus, termed $R_{50}$ runs) while T6 and A6 use 100 zones ($R_{100}$ runs).
Three additional, control models, N1--N3, have
set the magnetic field to zero. They vary $\chi$, but are
otherwise identical to the T1--T5 and
A1--A5 models. These purely gasdynamical cases are
intended to assist in identifying MHD effects on
the flows that might develop even when the field is initially weak.
In addition we have explored the degree to which flow behaviors depend on
the density contrast, $\chi$, enabling us to compare the current
simulations with our previous ones. All of the MHD models have assumed
$\chi = 10$, while the gasdynamical models, N1,N2 and N3 have
used $\chi = 10$, $40$ and $100$,
respectively. The duration of each run is indicated in Table~\ref{tbl-1}
as $t_{end}$.

The appendix evaluates ``convergence'' issues associated with the grid. 
To summarize quickly the appended discussion, we conclude that
the $R_{50}$ simulations are well converged by such global measures
as the quantities defined in Eqs.~\ref{mbull}--\ref{yav}.
The magnetic energy enhancement (Eq.~\ref{emag})
is reasonably well converged
in the transverse field cases and in the aligned field cases at
early times. However, the topology of the aligned field makes it
more sensitive to the details of eddy structures that form around the
bullet as it is destroyed, and those details depend on numerical resolution.
So, at late times the magnetic energy
enhancement is not converged in our aligned-field simulations. In association 
with this, it is clear that structural details of the bullets at
late times, $t/t_{bc}>>1$, do depend on resolution, especially since
the perturbations in the bullet structure that eventually lead to
its destruction come out of the mismatch between the bullet geometry and the
grid geometry in our simulations. Those aspects should
serve as reminders that calculations such as these are idealized efforts
to understand the physics of clouds interacting with their
environments and not intended to be used as predictors of
detailed structures.
That point is also made stronger by the recognition that bullet evolution,
particularly after $t/t_{bc} = 1$, depends physically on the initial
bullet structure, including the bullet geometry (see \cite{xustone})
and the density contrast (see discussion below).

Figs.~1, 2 and 3 contain images that
provide a summary spanning pretty well the
behaviors of the MHD models we have computed
before significant mass stripped from
the bullets begins to leave the
grid. Panels (a) and (b) in each of those figures
represent transverse field cases,
with  T2 ($\beta_0 = 4$) in (a) and T5 ($\beta_0 = 256$) in (b).
Panels (c) and (d) represent the analogous aligned field cases, with 
A2 in (c) and A5 in (d).
For each model two times are shown: $t/t_{bc} = 2$ and $6$. Fig.~1
illustrates evolutionary aspects of the log gas density, Fig.~2 shows the 
magnetic field lines and Fig.~3 the vorticity. In the
appendix, Figs.~a1 and a2 show log density and
field lines at $t/t_{bc} = 4/3$ and $4$ for models T5 and A5 as well as their
higher resolution representations, T6 and A6. For comparison,
Fig.~4 illustrates the log density distributions for the
gasdynamical models N1--N3 at $t/t_{bc} = 1$ and $2$. 

\subsection{Bullet Evolution: Aligned Field Cases}

It is convenient to begin our discussion with an outline of the
evolution of bullet structures and associated magnetic fields
in the cases with fields aligned to the direction of the bullet motion.
Those show less dramatic dynamical differences from the gasdynamical
bullets discussed in \cite{jonkant}, and closely resemble in some ways the
aligned field simulations of shocked MHD clouds described in \cite{maclow94}.
Several key points are clear from the figures listed above. First, in all
the aligned-field cases
the bullet shows signs of penetration by one or
more large R-T bubbles by $t/t_{bc} = 2$. There is little effect from the
magnetic field on the  bullet structure by this time. That impression
extends to the strongest field cases we considered;
namely, A2 $(\beta_0=4)$ and A1 $(\beta_0=1)$.
Indeed, the density distributions
for A2 and A5 at this time are very similar and also hardly
distinguishable from the analogous run with $B = 0$,
model N1 (see Figs.~1 and 4).
This comparison
is more quantitatively apparent in Fig.~5, which includes plots
of density, $\rho$, velocity as measured in the initial rest frame
of the bullet, $u_x$, gas pressure, $p_g$,
and magnetic pressure, $p_b$,
along a cut just above the $x$-axis for the
same situations as in Figs.~1, 2, and 3. The solid lines
correspond to $t/t_{bc} = 2$, while dotted lines illustrate quantities
at $t/t_{bc} = 6$. Furthermore, the vorticity structure around
the bullet is practically the same for both A2 and A5 (Fig.~3)
providing a very good indicator that the dynamics in the
two models are very similar at this time
(see, \eg, \cite{klein94}). Qualitatively
this vorticity structure is also the same as that found in
three-dimensional gasdynamical
simulations
of shocked clouds at comparable dynamical times (\cite{xustone}). The
vorticity associated with flow over the bullet is negative (clockwise).
We see in Fig.~3 that in the stronger field case, A2, some vorticity of the
opposite sign has been generated in the bullet wake in response to
magnetic tension there, however.

Comparing at $t/t_{bc} = 2$ model A5 in Fig.~1
with model N1 shown in Fig.~4, we see no detectable difference
in density distributions.
The field has not had any 
appreciable influence on the development
of R-T and K-H instabilities, since it is not strong enough anywhere
along the bullet boundary to suppress them directly. K-H instabilities will
be suppressed by the field along the boundary according to linear
theory, if the local Alfv\'en speed 
exceeds roughly the velocity difference across the boundary (\cite{chandra}). 
Since the local Mach number of the flow along the
bullet boundary is generally less than or about unity, the criterion for
the magnetic field removing K-H instabilities
is roughly $\beta \lsim 1$ along the bullet boundary.  Similarly
if the Alfv\'en wave crossing time through a bubble perturbation on the
bullet edge is less than the
``buoyancy rise time'', ${1}/{\sqrt{a\lambda}}$ where $a$ is the
acceleration and $\lambda$ is the length scale, the magnetic field will
inhibit R-T instabilities. Using Eq.~\ref{ub} (see below) to estimate the
acceleration of the bullet, we obtain a rough criterion for the
magnetic field to stabilize R-T instabilities; namely, $\beta \lsim \chi/M$. 
That leads in the present context again to $\beta \lsim 1$, since $\chi = M$.
Even in the
case A1 with $\beta_0 = 1$, the local $\beta$ along the bullet face
and sides is everywhere greater than unity. 
In fact, we do not find at any time in aligned field models A1--A6
that the magnetic field adjacent to the bullet is
ever strong enough that $\beta \leq 1$ (there is one region with $\beta<<1$
in the wake, as we will discuss below). However, even a relatively
weak magnetic field can play a significant role in small scale flow dynamics,
as pointed out by others (\eg, \cite{cattvain}; \cite{franetal}) and
as we shall explain further on.

By $t/t_{bc} = 2$ the states of K-H and R-T
instabilities are, on the other hand, rather
strongly influenced by the density contrast choice, $\chi = 10$. 
The other panels in Fig.~4 illustrate how, as the density contrast
is increased, the penetration of the R-T bubbles into the bullet
is enhanced for fixed $t/t_{bc}$.
While, in the limit of large $\chi$, the linear growth times
on a given length for both K-H and R-T instabilities 
scale directly with $t_{bc}$, both
growth rates include an additional term 
that increases with $\chi$ when it is finite.
That tendency is consistent with the behaviors illustrated in Fig.~4.
As a further test, we can compare run N3 ($\chi = 100$) with
very similar gasdynamic simulations reported in \cite{jonkant}.
Those earlier simulations were carried out using a PPM gasdynamic code
and were designed to study dynamical feedback on bullets
from cosmic-rays accelerated in shocks associated with the bullet.
However,
model number 1 in that paper omitted cosmic-rays. Their Fig.~4
illustrates density structure at $t/t_{bc} = 2$, 
showing a good
correspondence between the cloud structures in the two simulations, except
that our current calculations show less evidence of small scale K-H
structures along the bullet boundary.
Even when ${\bmit B} = 0$ the MHD TVD code does suppress
the smallest scale K-H instabilities
in comparison with the PPM code, because it spreads the very strong 
contact shear layer of
the bullet edge over several more zones.  K-H instabilities are suppressed  
on scales less than the thickness of the shear layer.

As a direct consequence of the penetration of the R-T bubbles, the 
bullet begins appreciably to expand laterally by $t/t_{bc}\sim 2$. 
For all the aligned field cases,
the mass weighted bullet height, $\bar h_{bul}$, has expanded to roughly
five times or more its initial value by $t/t_{bc} = 6$ (see Fig.~6).
This characteristic behavior was also noted by others with regard to 
gasdynamic simulations. As illustrated in Fig.~a1,
before $t/t_{bc} = 4$, our bullets develop a roughly ``C''-shaped morphology 
on the computed half-plane in all of the 
aligned field cases, A1-A5. (Including the reflected portion below
the computed space, the
cloud shape would be an ``E''.)\footnotemark~ That ``C'' continues to open up
and thin itself even until the bullet moves off the grid. As the figures
show bullet mass is stripped and carried into the wake, eroding the bullet
body. 

\footnotetext{As indicated earlier, the specifics of bullet morphology
depend on a number of details, both physical and numerical.  So the ``C''
shape more generally represents the fact that some number of dense
R-T fingers will protrude
in a forward direction, while light R-T bubbles will push into and expand
the bullet body.}

In response to drag forces the bullet is accelerated towards
a terminal velocity, $u_{b0}$, as measured
in the original reference frame; that
is, the bullet or its fragments should come to rest in the surrounding flow.
\cite{klein94} derived a simple theoretical model for the acceleration of
a cloud based on ram pressure,  taking into account the lateral 
expansion. The drag force actually comes from the difference in the
total pressure across the bullet. Assuming highly supersonic motion,
we can estimate in the gasdynamic case from Bernoulli's equation 
the pressure difference on the symmetry axis to be
$\Delta p \approx (4/5) \rho_a (u_x - u_{b0})^2$,
where $u_x(t)$ is the instantaneous speed of the bullet with respect
to its initial reference frame and we have set
$\gamma = 5/3$. Applied across the full bullet this
leads to the usual expression for its acceleration, 
\begin{equation}
\frac{d\bar u_b}{dt} = \frac{3}{4}
\frac{C_d}{t_{bc}}\frac{(u_x - u_{b0})^2}{\sqrt{\chi} u_{b0}}\frac{r^2(t)}{R^2},
\label{accel}
\end{equation}
where $C_d$ is a drag coefficient that absorbs our ignorance of
details of the pressure distribution and $r(t)$ is the effective bullet
radius as a function of time. Borrowing the notion in \cite{klein94}
that the bullet
expansion begins only after
$t/t_{bc} = 1$, we write $r^2(t) \sim R^2[1~+~C_e ({t}/{t_{bc}})^2]$
where we can term $C_e$ an expansion coefficient (see also \cite{zahn92},
\cite{macz94}). If $C_e > 0$,
drag on the bullet increases with time, enhancing its acceleration.
In an MHD flow we have to account for
magnetic pressure in Eq.~\ref{accel}.
For aligned cases, however, the magnetic pressure does not enter
into the Bernoulli equation on axis and the magnetic pressure
is not significant anyway. Thus we expect the bullet motion to
behave pretty much as in the gasdynamic case. We will comment later
on modifications to Eq.~\ref{accel} appropriate to transverse
field cases.
Eq.~\ref{accel} can be integrated to give
\begin{equation}
\bar u_b = u_{b0} \left( 1~-~{1\over{1~+~{3\over4}~
\frac{C_d}{\sqrt{\chi}}~\left(\case{t}{t_{bc}}\right)~
\left[1~+~C_e\left(\case{t}{t_{bc}}\right)^2\right]}}\right).
\label{ub}
\end{equation}
All the aligned field cases can be fit well with a
drag coefficient, $C_d \approx 1$, and
an expansion coefficient, $C_e \approx 0.05$. There is a weak trend visible
in Fig.~6 for the
acceleration of the bullets to be faster when the field is smaller, 
corresponding to a value of $C_e$ that depends inversely on $\beta_0$. That
just reflects the fact that stronger aligned fields do resist lateral
expansion, keeping the bullet cross section somewhat smaller.

We expect the magnetic fields themselves to respond differently
to the motion of the bullet for the two field geometries we have used. 
That is apparent even by $t/t_{bc} = 1$
as shown in Fig.~a2.
The evolution of the
aligned field is similar to that found by \cite{maclow94} for shocked
clouds with aligned fields. 
To the front of the bullet, field lines that initially
pass through the bullet are swept 
and then ``folded'' over the top of the bullet in a configuration
that is unstable to
the resistive tearing mode. That instability leads to ``magnetic islands''
within and behind the bullet as seen at $t/t_{bc} = 1$ in Fig.~a2 c and d
or at $t/t_{bc} = 2$ in Fig.~2 c and d.
The compact magnetic island closest to the axis coincides
with the strong vortex at the rear
of the bullet (Fig.~3). Magnetic flux initially formed into these
islands is mostly
annihilated  by about $t/t_{bc} = 4$ (or more properly ``expelled'', since 
the total magnetic
flux through the computational box does not change over time for the aligned
field configuration). Magnetic reconnection also takes place
inside the bullet in consequence of the circulation developing after
the bullet shock has exited (see Figs.~2, 3, and a2).
That significantly reduces the field strength within the bullet over time.
Through these reconnection events, magnetic flux is separated into
two elements: that which passes around the bullet without reversal and
that which passes through the bullet. The latter flux element continues
to be involved with vortices around the bullet and subjected to reversals
and reconnection.

Before the magnetic flux separation, there is
a thin region of strongly concentrated field formed along the axis
behind the bullet, analogous to the ``flux rope'' discussed for shocked clouds
in some detail by \cite{maclow94}. In our situation, as in theirs, this
feature forms as a consequence of compression of field into the
low pressure wake behind the bullet, followed by field-line stretching.
In our case those field lines are temporarily anchored in the bullet on one
end and drawn out to the rear by the expanding rarefaction wave (see Figs.~a1 and a2).
Just as for the shocked cloud case considered by
\cite{maclow94}, the magnetic field in this region can become locally
dominant with $\beta \leq 10^{-1}$. However, as Fig.~2 shows
clearly, the field configuration on the edge of the flux rope
is susceptible to tearing mode instabilities,
so that the rear flux rope disappears as
part of the flux separation event.
We do {\it not}
see in any of our aligned field simulations that $p_b \sim \rho_a u^2_b$,
nor that this flux rope plays a significant dynamical role in the
evolution of the bullet. However, this feature is the dominant source
of enhanced magnetic field energy before its flux is expelled (see Fig.~6).
The normalized magnetic energy enhancement,
$\Delta E_{mag}$, (Eq.~\ref{emag}) peaks 
at values between $10$ and $100$ near $t/t_{bc} \sim 1.5$ (see Fig.~6). 
The peak value in model A1 ($\beta_0 = 1$,
$\Delta E_{mag} = 18$) is about half that
reported by \cite{maclow94} for their shocked cloud
during this phase for the same numerical
resolution across the cloud and the same $\beta_0$.
Since our post-bullet flux rope is
significantly thinner than theirs, the comparison seems very good. Our
higher resolution, $R_{100}$, weak field run, A6,
differs in $\Delta E_{mag}$ by 
$\lsim 20$\% from the analogous low resolution, $R_{50}$,
A5 run for $t/t_{bc} < 4$. So, through
this stage the magnetic energy behavior seems well converged.

For the aligned field geometry there is very little compression of the field,
so from the start we should expect that field line stretching would
be the primary contributor to magnetic field enhancement.
Along the boundary of the bullet the field remains
weak before about $t/t_{bc} = 4$, even in the
$\beta_0 = 1$ case, because it
is subject to reconnection that shortens the field lines.
However, that reconnection
leads directly to the flux separation mentioned, and
through that the field topology above the bullet returns
to something resembling its initial form; namely, field lines
pass directly around the bullet from front to back without folding.
As the bullet body expands laterally,
lines above it are stretched significantly
(see Fig.~2), but are no longer subject to reconnection. 

A relatively thin flux tube especially on the rear perimeter of the bullet envelops 
the now-distended
cloud, extending into the wake. Almost independently of the initial
$\beta_0$ or the numerical resolution, the minimum local $\beta ~{\rm is}~\sim 10$
in this structure.  After the flux separation event these field lines
do not penetrate the strong vortex at the rear of the bullet.
On the other hand, the flux that now passes through the bullet is
drawn into the big R-T bubble and an associated vortex pair.
Through reconnection, however,
those field lines divide into magnetic islands (for the weaker initial
field cases) that are annihilated and flux that penetrates
directly through the bullet.
Some flux passing through the bullet is drawn
into the strong vortex at the rear
and base of the bullet, leading to another region of strong magnetic flux 
on its perimeter.
Even though the magnetic pressure in these flux
tubes is never dominant, the field can still play a major role in reducing
the vorticity and leading the
flow to become more nearly laminar and less disruptive.
That dynamical behavior was shown recently in
high resolution simulations of the MHD
K-H instability (\cite{franetal}). There
it was found in K-H unstable MHD flows that even when $\beta \gsim 30$
the magnetic field, acting as catalyst, realigns the flow into a stable,
broad and laminar shear layer. The field acts as a catalyst in the
sense that kinetic energy is temporarily stored in the field
so that locally the magnetic tension is at least significant if not
dominant. That stored energy is released again during reconnection, but
one result of the reconnection is that the velocity and magnetic fields
are more closely aligned; that is, the magnitude of the cross helicity 
is increased. The smoothing and spreading of the
flow should significantly reduce stripping from the bullet. This conclusion
is consistent with that reached by \cite{maclow94}, that the magnetic
field reduces the intensity of vortices around their clouds and, thus,
increases the cloud survivability.
The presence of the magnetic field is clearly felt by the bullets in
our aligned field simulations.
Clouds in the stronger field cases are less distended (Figs.~1, 2, and 6)
and there is an apparently
stable density concentration at the top of the bullet ``C''. Examination of
the forces applied there shows that the density is mostly confined
by ram pressure, but that magnetic pressure and tension contribute
at the 10\% level, as anticipated from earlier discussion.
At late times in these cases material no longer seems to be
stripped from the bullets (see Fig.~6), consistent with our comments 
above. For the stronger field cases, there
is less thinning of the main bullet body, as well.

Magnetic energy enhancement in the aligned field models begins to rise
again after $t/t_{bc} \sim 4$, mostly in response to the development of 
flux tubes around the bullet and in the associated vortices. Our 
simulations are not able to capture this final rise completely, because
the flux tube fine structures are still resolution dependent
(see Figs.~a2 and a4) and because  significant
magnetic structures begin to leave the grid
after $t/t_{bc} \sim 7$. It appears likely,
however, that the excess magnetic energy becomes at least comparable in
this stage to the peak value noted earlier. \cite{maclow94}
similarly emphasized the lack of convergence in the magnetic field
within their simulations. They also used an aligned field
geometry, so the issues responsible were analogous.

Although the details are fairly complex, the summary of aligned-field
bullet evolution is straightforward. Magnetic fields are initially swept 
over the bullet, stretched and folded there.
Eventually reconnection separates the
field into flux passing directly through the bullet and flux passing directly
around it. Except within a thin flux tube formed temporarily along
the symmetry axis, there is no place that the magnetic field
becomes energetically dominant; that is, almost everywhere $\beta >> 1$.
In the meantime the disruptive, ram pressure force that applies
to gasdynamical bullets (and gasdynamical shocked clouds) causes
the initial cloud to become distended
and stripped. Magnetic field stretched over the top of
the bullet can have a significant stabilizing influence that should
prolong bullet coherence even though
magnetic energy never becomes dominant.

\subsection{Bullet Evolution: Transverse Field Cases}

For the transverse cases where the initial field
crosses the bullet's path, field lines are
also swept and stretched around the bullet. However in this case
the field lines do not reconnect around the bullet, except in a ``magnetotail''
along the axis, so that a region of high magnetic
pressure develops on the bullet nose as shown in Fig.~2. Even in the
T5 case with $\beta_0 = 256$, the field is strong enough that
$\beta \sim 5$ on the
bullet nose by $t/t_{bc} = 2$. For all the other transverse field
cases we computed, $\beta \lsim 1$
along the nose of the bullet by this time. In
models T1 and T2, $\beta < 10^{-1}$ here and $p_b \sim \rho_a u^2_b$ directly
on the leading edge of the bullet.
Except in T5, $\beta \lsim 10^{-1}$ along the top and towards the rear of
the bullet boundary, partly because the gas pressure is low, but mostly
because field lines have been greatly stretched. That characterization
of the field lines is obvious in Fig.~2 or Fig.~a2.  Even though field
lines are compressed by the bow shock, it is stretching, not
compression, that is important to the evolution of the magnetic field
and to its eventual dynamical role. That supports the
expectations expressed by \cite{jonkant}, based on passive fields. One
can see in Fig.~6 that the normalized magnetic energy enhancement
for the transverse geometry grows to more than $10^3$
for the weakest field cases and more
than $10^2$ even for the strongest field cases.
Since the volume containing highly compressed plasma is never more
than a few times the initial bullet volume, the maximum field
enhancement through compression would be $\lsim 10~E_{Bob}$,
emphasizing the importance of stretching. We note
that the magnetic energy in the transverse cases seems fairly well
converged through the full duration of the simulations and better 
than in the aligned field geometry. That makes sense, since the
transverse fields are not so closely tied to vortical flows behind
the bullet.

We can write down a crude model for the enhancement of the magnetic field
energy in the transverse cases that seems to qualitatively account for
what is seen. For ideal MHD the magnetic induction equation can be expressed 
in a form
\begin{equation}
\frac{d\ln{B/\rho}}{d t} = {\frac{{\bmit B}\cdot
[({\bmit B}\cdot{\bmsy\nabla}){\bmit u}]}{B^2}},
\label{dbdt}
\end{equation}
where $d~/dt$ is the convective derivative.
At the stagnation point on the bullet nose we can estimate
${\bmsy\nabla}\cdot{\bmit u} \lsim \case{1}{4} u_{b0}/R$
and ${\bmit B}\cdot [({\bmit B}\cdot{\bmsy\nabla}){\bmit u}]
 \sim B^2 u_{b0}/R$, so that
\begin{equation}
\frac{d\ln{B^2}}{d (t/t_{bc})} \sim \alpha \chi^{\case{1}{2}},
\label{bgrowth}
\end{equation}
where $\alpha \sim 1$ is a ``fudge factor'' accounting for 
various details. If this field growth 
occurs within a ``shield'' of volume $V_{shield} \sim R^2$, we can
estimate the associated magnetic energy enhancement to be
\begin{equation}
\Delta E_{mag} \sim few\times {\exp\left(\alpha \chi^{\case{1}{2}} \case{t}{t_{bc}}
\right)}.
\label{emodel}
\end{equation}
That growth seems consistent with the results shown in Fig.~6, for
$t/t_{bc} \lsim 2$.
Since the field on the bullet nose saturates with $p_b \sim \rho_a u_{b0}^2$,
we expect a maximum for $\Delta E_{mag} \sim few\times\beta_0 M^2$. That is
also roughly consistent with the results in Fig.~6.
The early magnetic energy enhancement is similar for all the values
of $\beta_0$, but it saturates sooner and the dynamical influence of
the field is sooner when $\beta_0$ is smaller.

The field lines draped over the bullet are also drawn down near the
symmetry axis into
the bullet wake, producing a region that
resembles the flux rope seen in the aligned field cases. Again, this is
a region of $\beta << 1$,  but primarily because
the gas pressure is very low from the initial evacuation of
this region. It is always the case in our simulations
that $p_b << \rho_a u^2_b$
within this feature. In addition, the $y = 0$ boundary of our
grid separates field lines of opposite direction; \ie, there is a
current sheet there, much as in the earth's magnetotail. That is
also unstable to tearing mode instabilities that limit fields
in this region. 

Like the earth's magnetosphere does, the penetration of the bullet
through a ``quasi-transverse'' field should produce on the
bullet nose an induced
electric field perpendicular to both the magnetic field and the
bullet velocity. In this case that projects out of the computational
plane. If the third dimension of the bullet $\sim R$, then the
induced potential across the bullet would be $\Phi \sim u_{b0} B R / c$.
Taking numbers that might be appropriate for knots in a young SNR
like Cas A ($B \sim 10^{-4}$ Gauss, $u_{b0} \gsim 10^{-3}$ c, $R \sim 10^{16}$ cm) (\eg,~ \cite{ander94}),
we find that $\Phi \gsim 10^{11}$ Volt is possible. Supposing that 
reconnection within
the bullet magnetotail can generate electric fields aligned with
the magnetic fields, then these regions might be important sites for
nonthermal particle acceleration.

For the weak transverse field cases, the bullet dynamics initially
resembles
the aligned field and gasdynamical cases to the first approximation.
The R-T bubbles seen in those other situations form here, too. In fact
at $t/t_{bc} = 2$ there are no obvious differences in either bullet
morphology or flow dynamics between models A5 and T5,
for example (see Figs.~1 and 3).
As time progresses, however, the bullet in all transverse field cases becomes
significantly influenced by the field. Even though the same
``C''-shaped bullet morphology is seen in the two weakest field
cases, T4 and T5, field stretched over
the top and back of the ``C''-shaped cloud develops magnetic pressures
significantly greater than the gas pressure, so that $\beta < 1$ along
the bullet edge. This condition protects the bullet from 
further disruption. In model T5 that happens quite late, so as late as 
$t/t_{bc} = 6$ the bullet morphology is pretty similar to A5. 
However, even in model T4 ($\beta_0 = 64$) the field becomes dynamically
strong ($\beta \sim 1$ along the bullet nose just after $t/t_{bc} = 2$).
That case still expands laterally, but not so fast as in the cases with
negligible magnetic influence (see Fig.~6). For all the other
transverse cases with $\beta_0 \leq 16$,
the Maxwell stresses grow large enough to completely inhibit
lateral expansion of the bullet.
The bullets are then enshrouded by a strong magnetic
shield. For cases involving an initially weak field, the bullet is partially
disrupted, but eventually the shield protects the R-T fingers from
further erosion. The bullet embedded in a stronger field is compressed,
but then, as the field confines it,  develops a streamlined profile
and is not strongly eroded.

Remarkably, Fig.~6 shows that the acceleration of the transverse-field
bullet increases with the strength of the magnetic field. This trend is
reversed from that of aligned-field bullets. There we have observed that
the tendency of stronger field to resist lateral expansion reduces
the evolution to a greater cross section. 
It was that increased cross section and the augmented drag that were 
modified by the field. For the transverse field
models, lateral expansion can be halted entirely, yet the drag force
clearly is enhanced with the stronger field. This seeming paradox
is easily explained by accounting properly for the role of the
magnetic pressure in the drag. Whereas the aligned field did not
contribute directly to the force across the bullet, the transverse
field does. The MHD Bernoulli equation on the symmetry axis gives
the result in the highly supersonic limit that 
$\Delta p = \Delta (p_g + p_b) \approx
(4/5) \rho_a (u_x - u_{b0})^2 + (1/5)p_b$,
where $p_b=\case{1}{2} B^2_y$ on the nose of the bullet in this case.
As already noted, $p_b$ increases over the first few crushing times and can
become comparable to $\rho_a u^2_b$ in that region. 
Although the expansion coefficient $C_e = 0$ in Eqs.~\ref{accel} and \ref{ub}
the drag coefficient, $C_d$, is an effectively increasing function of time
and that enhances the bullet's acceleration.
We find, for example,
that Eq.~\ref{ub} gives a good fit to the motion of the strongest
field case, T1, by setting $C_e = 0$, replacing
$C_d$ with $C_d(1 + 0.01 t/t_{bc})$,
and using the same $C_d = 1.0$ that we have used for aligned field models.
That corresponds to an increase in drag of about 5\% at the end of
the simulation.

Thus, the picture that develops for the transverse field bullets is rather
different from and considerably simpler than the ones for aligned field.
Magnetic field is swept around
the bullet to form a protective shield. The magnetic pressure
becomes comparable to the ram pressure through the bullet bow shock; thus,
the magnetic energy is enhanced by an amount
approaching $\sim M^2 \beta_0 p_{b0} V_{shield}$.
Again taking that to
be roughly the volume of the bullet, we recover the observed
magnetic energy enhancements with
the expression $\Delta E_{mag} \sim\beta_0 M^2$.
Our bullets have a cylindrical form, so that field lines cannot be swept
around the sides of the bullet in the $z$-direction.
Such sweeping may reduce the magnetic
field enhancement somewhat in a three-dimensional bullet.
But we still expect the
same qualitative behavior, since much of the field that forms into
the shield comes from the field lines that penetrate into the ``skin'' of
the bullet, and since irregularities in the spherical bullet would capture
field lines much like the cylindrical one does.

The brief summary of transverse bullet evolution is the following.
Even when the initial field is of modest strength measured in terms of
the ratio of magnetic pressure to background plasma pressure, the stretched magnetic
field effectively confines the bullet and prevents its fragmentation. Further,
the magnetic pressure applied to the nose of the bullet increases the
rate at which the bullet is brought to rest with respect to the ambient
medium; more effectively, in fact, than the lateral expansion that
accompanies the acceleration of gasdynamical bullets.

\section{Summary and Conclusions}

We have carried out an extensive set of two-dimensional
MHD simulations exploring
the role of magnetic field in the dynamics of supersonic clumps of
plasma. We have examined the influence of both field strength and
orientation on the problem. Of those two characteristics field
orientation is far more important. Even a very modest ambient magnetic
field that crosses the path of the bullet tends to be amplified
by field line stretching around the bullet until the Maxwell stresses
become comparable to the ram pressure associated with the bullet motion.
A field that is aligned with the bullet motion, on the other hand, 
develops reconnection-prone topologies that shorten the stretched field
and release the excess energy it contains. The field is also swept
around the bullet in this geometry and can temporarily become moderately
strong. However, the energy in the field is not enhanced so
much for this geometry as for the transverse geometry.  Rather, as a 
consequence of reconnection, there is a transformation and relaxation
of the field over a few bullet crushing times. Some magnetic
flux passes directly through the bullet, where it can become entrained
in vortices and amplified around the vortex perimeter. The remaining 
flux passes directly around the bullet, where
it can act to resist lateral expansion as it is stretched by
that expansion. In this geometry, however, 
the Maxwell stresses on the bullet never approach the ram pressure level.
Even so, a field of even moderate initial strength becomes
strong enough to help realign the flow around the bullet into a smoother,
more laminar form that reduces the tendency of the  bullet to fragment.

Both of the above field geometries are highly idealized, and more
generally the bullet would encounter a field at some oblique angle
to its motion. In that situation, we will want to know which of the two
special cases is more relevant. The most important detail should be if, as the
field lines are swept over the bullet, they are folded over the top
as well, thus leading to reconnection. Conceptually that depends on 
the relative rates at
which the field lines are swept past the bullet body, on the one hand, and
at which the ``foot-points'' of the field lines along the bow shock
move past the bullet, on the other hand. If the field-line bow-shock
foot-point moves faster (as it must for transverse field cases), the
lines are not folded over the top, but stretched directly to the
bow shock. If, however, the field-line bow-shock
foot-point moves slower (it is at rest for aligned field cases),
then the field lines are folded over inside the bow shock on one
side of the bullet and likely to reconnect. 
We can derive an approximate expression for the condition that 
the field foot-point moves downstream faster than the flow around the
bullet, if we take the Mach bow shock to be a simple cone with half angle
$\arctan(1/\sqrt{M^2 - 1})$ and suppose that the flow speed around the
bullet is $u = \delta u_{b_0}$.  
Defining the motion of the foot-point as the translation of the
intersection between the field line and the bow shock,
this leads to the constraint
\begin{equation}
\tan{\theta} \gsim {\frac{\delta}{(1~-~\delta)\sqrt{M^2 - 1}}},
\label{coneq1}
\end{equation}
where $\theta$ is the angle between the bullet motion and the ambient
magnetic field.
In the limit $M>>1$, with $\delta \sim 1/2$, this becomes  
\begin{equation}
\tan{\theta} \gsim {\frac{1}{M}}.
\label{coneq2}
\end{equation}

So, we conclude that for supersonic bullets most field directions will
lead to behaviors similar to the transverse field cases, while as the
motion becomes transsonic the dividing line
would be closer to 45\arcdeg. Except in circumstances with preferential
alignments between the motion and the field, that would lead
to the further conclusion that
even a weak magnetic field will have a substantial impact on the
evolution and dynamics of supersonic clumps of ionized, conducting 
gas. In young supernova remnants, polarized radio synchrotron emission
indicates that there is a net radial direction to the magnetic field inside
the remnants (\cite{milne}).
This might suggest that bullets in young remnants would
generally encounter a radial field and, hence, one nearly aligned with
their motions. That view is somewhat simplistic, however, because the
same radio observations also show
a very small net polarization (\cite{ander95})
indicating that, to first order, the field is disordered. That, in turn,
argues that a wide range of field orientations may be encountered by
small projectiles.

In either limiting field geometry and presumably those in between, field
line stretching is the dominant process for magnetic field amplification.
That supports the conclusions reached
by \cite{jonkan93} and \cite{jonkant} that nonthermal radio emission
associated with supersonic clumps in supernova remnants, for example,
is likely to be largely controlled by the generation of stretched
magnetic fields around the perimeters of the clumps, rather than primarily
highlighting the bow shock where field is mostly enhanced by compression.
We note that the structures of the bow shock
and the bullet boundary may even superficially
resemble each other in observations, so that morphology alone can
be misleading. The physical difference is
important, however, since the bow shock is probably the site of local
particle acceleration, while on the bullet perimeter one sees
primarily energetic particles that come from some other site (possibly
including the bow shock, of course). Our estimates of the field strengths
expected in the two situations are very
different, however. So, observational estimates of the local magnetic
field (based on equipartition, for example) would possibly lead us to 
very different conclusions about the local conditions.

In addition, since the magnetic pressure on the nose of the bullet may
become comparable to the ram pressure and hence the total pressure
behind the bow shock, the gas pressure there could be substantially lower than
that in a gasdynamical bullet. That means, as well, that the temperature
in the region on the nose of the bullet would be lower than
that predicted in the gasdynamical case. That detail can alter
expectations of the thermal emission, including X-rays and UV-IR
lines.

In summary, the role of a magnetic field in the evolution and in the
appearance of supersonic clumps is very important, even if the
magnetic field is nominally not strong in the ambient medium. The importance
comes because magnetic field lines can be stretched and amplified if they
become draped around the bullet perimeter. That effect seems especially 
strong when the field lines are ``quasi-transverse'' to the motion, a
concept that depends on the speed of the bullet, but seems to
include most directions. Additional amplification within vortices
associated with the destruction of the bullets can also occur, but
seems to be less important. If the fields are quasi-transverse, then
they can effectively confine the bullet and prevent its disruption.
The same amplified fields may be important to emissions used to
analyze the bullets. They can control the radio synchrotron emission
expected and possibly become strong enough to alter the local thermodynamics
of the gas and influence thermal emissions, as well.

\acknowledgments

We are grateful to Adam Frank, Byung-Il Jun, Bob Lysak and Larry Rudnick for
fruitful discussions and to Mordecai-Mark Mac Low for helpful comments on
the manuscript.
At the University of Minnesota this work was supported in part by
NSF (AST-9318959), NASA (NAGW-2548), and the Minnesota Supercomputer
Institute.
At Chungnam National University this work was supported in part by
the Basic Science Research Institute Program, Korean Ministry of
Education 1995, Project No.~BSRI-95-5408.

\appendix
\section{Numerical Grid Issues}

We have carried out several experiments to understand the influence of
finite numerical resolution on our simulations. There are two issues.
First, we are nominally
simulating ideal MHD, but depend on the existence of a small, albeit 
finite, dissipation on the grid cell scale to simulate the physical
viscous and resistive dissipation that
certainly takes place on very small scales.
The existence of the numerical dissipation is necessary, for
example, to allow shocks to form and magnetic reconnection to occur.
There is fairly good evidence that conservative monotonic
schemes such as ours do a good job of approximately representing
physical viscous and resistive dissipative processes presumed to exist
on scales somewhat smaller than the grid (\eg, \cite{porwood94}).
For the astrophysical environments being simulated, the 
dissipative scales are probably very much
smaller than those that can be modeled
directly, however. Thus, the numerical solutions, if they are complex, will 
generally differ to some degree from the asymptotic ``physical''
solution  on the smallest scales after long periods of time.
One hopes, on larger scales and over moderate periods of
time, that the behaviors will be converged
for important characteristics. Our code is second-order
accurate in smooth flows, so that 
the effective Reynolds numbers
increase with length, $l$, approximately as $R_e\sim K (l/\Delta x)^2$
where $\Delta x$ is the size of a computational zone (\cite{ryujonf}).
For moderately strong magnetic fields ($\beta \sim 2$), $K \approx 0.5$
in our code.  For much weaker
fields the dissipation is several times less on a given scale.
In any case, we expect that
dissipation would be confined to scales of only a few zones and that
{\it inertial} structures formed within
scales of less than $10-20$ zones would
have histories that are resolution-dependent.
The second, related issue has to do with the fact that the late history
of a bullet depends on the nonlinear evolution of instabilities
that in our simulations form off of the initial surface of the bullet. Consequently that
history can depend to some degree on the exact structure of the initial
perturbation in a way that extends beyond the size scale of the perturbation.

Most of our simulations (T1--T5 and A1--A5)
have been conducted using a grid placing
50 zones across the bullet radius ($R_{50}$) and with the bullet
centered in the middle of a zone along the bottom axis $(y=0)$.
To evaluate the importance of the zone size on our calculations,
we have carried out simulations
identical to T5 and A5 except that 100 zones spanned
the bullet radius ($R_{100}$; T6 and A6).
In addition, we have performed a number of test runs with varying
resolution using this code but
setting the magnetic field strength to zero.
In some of those, we have simulated the impact
of a Mach 10 shock on the cloud.  The intent there was to repeat the
resolution tests shown in \cite{maclow94}. Since they used 
cylindrical geometry, and we have used Cartesian geometry, an exact
comparison is not meaningful.  But we conclude that our MHD code 
produces resolution dependencies very comparable to those they displayed
as computed from their gasdynamical code.

Some simple illustrations make plain the limitations imposed by our grid.
Fig.~a1 compares the density structures computed for models
T5 and T6 (top panels) as well as A5 and A6 (bottom panels). Results
in each case are shown at two times, $t/t_{bc} = {4}/{3}$ and $ 4$,
with the earlier time displayed above the later time.
Fig.~a2 displays the magnetic field structures for the
same cases. $R_{50}$ runs, T5 and A5, are shown on the left and
$R_{100}$ runs, T6 and A6, on the right.
These have the weakest initial fields of any of
the simulations we have run in this
study. They are, thus, the cases where we would expect the greatest dependence on
numerical resolution, since a strong magnetic field has a tendency to
produce smoother flows and more organized field structures.
At $t/t_{bc} = {4}/{3}$
the paired runs agree almost exactly for both field
configurations. At $t/t_{bc} = 4$ the structural agreement
is qualitatively consistent, but
there are clear differences. Especially in the magnetic
field lines shown in Fig.~a2, there is additional fine structure
in the higher resolution runs.  Also the field
along the leading edge of the bullet is somewhat stronger in the
higher resolution runs. 
Both of these features are anticipated, since they come directly from the
reduced dissipation scale in the $R_{100}$ runs.
Most apparent, however, is that the instabilities
that begin to destroy the bullet have a somewhat different nonlinear
development; \ie, the shapes of the bullets are beginning to differ. 

As it turns out, that result comes at least as much from the fact that
the form of the initial perturbations
on the bullet is slightly different as a consequence of the different
resolutions as from the differences in dissipation scales.
The reason is because our initial
perturbation results from the mismatch between a circular cross
section and a Cartesian grid. Thus, either a change in the number of
zones spanning the bullet or a shift in the bullet center will
alter the locations of those irregularities. To make that point 
clear, we compare in Fig.~a3
the density distributions for two $R_{50}$ runs at $t/t_{bc} = 3$ and $ 5 $.
The one shown on the
left is A5, while the one on the right is identical to A5 {\it except for
one small detail}; namely, that
the initial bullet center is shifted ${1}/{2}$ zone to the
right from a zone center to a zone edge along the bottom grid boundary. 
The differences between these two
runs are almost as great as those between the two different resolutions,
and, in particular, the degree of the changes in bullet shape is comparable.
This emphasizes that, once the instability becomes nonlinear,
the precise structure becomes sensitive to exact
details of the initial conditions. 

Considering these comparisons, one
should concentrate on more global characteristics than on 
specific structural features in simulations
like those in this paper. In that global sense,
the model comparisons are much closer. To demonstrate this, Fig.~a4
compares the evolution of the bullet mass, speed, and height as well
as the normalized magnetic energy enhancement in the computational domain
for each resolution pair (T5,T6 and A5,A6). In each pairing, these important
dynamical quantities agree very well. For example, the magnetic energy 
enhancement in the two transverse field runs approaches $\sim 10^3$
by $t/t_{bc} = 4$. The two runs
agree within about 10\% on that figure. In the aligned field
pair, the enhancement is much smaller and $\Delta E_{mag}$ peaks
between 50 and 60 for the two resolutions with about $20\%$
more enhancement in the $R_{100}$ run.
Those differences reflect, on the one hand, that for the transverse
field cases the magnetic field is more effectively stretched, but,
on the other hand, that for the aligned field cases the field
interacts much more with vortices. Vortex structure and magnetic
field behavior there are fairly sensitive to numerical resolution.
Overall, it is clear that run  pairs are largely consistent with 
each other. Slightly less stripped bullet material
has left the grid in each of the high resolution runs at the
last time shown, but is less than 1\% of the bullet
mass, in any case.

\clearpage
 
\begin{deluxetable}{ccccccc}
\footnotesize
\tablecaption{Summary of Simulations Reported \label{tbl-1}}
\tablehead{
\colhead{Model\tablenotemark{a,b}} & \colhead{Resolution\tablenotemark{c}}
&\colhead{$\beta_0$\tablenotemark{d}}  &
\colhead{$M_A$\tablenotemark{d}} & \colhead{$\chi$} &  
\colhead{$t_{bc}$\tablenotemark{e}}  & \colhead{$t_{end}$}
} 
\startdata
T1 (A1) &$R_{50}$ &1 & 9.13&10 &0.6 &5 $t_{bc}$\nl
T2 (A2) &$R_{50}$ &4 & 18.3 &10 &0.6 &8 $t_{bc}$\nl
T3 (A3) &$R_{50}$ &16 & 36.5  &10 &0.6  &8 $t_{bc}$\nl
T4 (A4) &$R_{50}$ &64 & 73.0 &10 &0.6 &8 $t_{bc}$\nl
T5 (A5) &$R_{50}$ & 256 & 146 &10 &0.6 &8 $t_{bc}$\nl
T6 (A6) &$R_{100}$ &256 & 146  &10 &0.6 &4 $t_{bc}$\nl
N1 &$R_{50}$ &$\infty$ & $\infty$ &10 &0.6 &2 $t_{bc}$\nl
N2 &$R_{50}$ &$\infty$ & $\infty$ &40 &1.2 &2 $t_{bc}$\nl
N3 &$R_{50}$ &$\infty$ & $\infty$ &100 &2.0 &2 $t_{bc}$\nl
 
\enddata

\tablenotetext{a}{All models have used $\gamma$ = 5/3,
$M = 10$, $u_{b0} = 10$, and $c_{sa} = 1$.}
\tablenotetext{b}{Models with fields transverse to the bullet
motion are designated by the letter ``T'', fields aligned with the
motion by the letter ``A'' and those with no magnetic field by the letter
``N''. All the ``T'' and ``A'' models have been done in pairs, with
otherwise identical characteristics.}
\tablenotetext{c}{Code indicates the number of computational zones
spanning the initial bullet radius. That radius corresponds to
$0.97656$ in the full computational domain which has dimensions
$x\times y = 10\times 5$.}
\tablenotetext{d}{$\beta_0 = p_g/p_b = (2/\gamma)
(M_A/M)^2$, defined in terms of background plasma values.}
\tablenotetext{e}{Values of the bullet crushing time used in the discussion.
For $\chi = 10$ and $40$ these times have been rounded down slightly for
ease of matching to simulation dump times.}
 
\end{deluxetable}

\clearpage

\begin{center}
{\bf FIGURE CAPTIONS}
\end{center}

\begin{description}

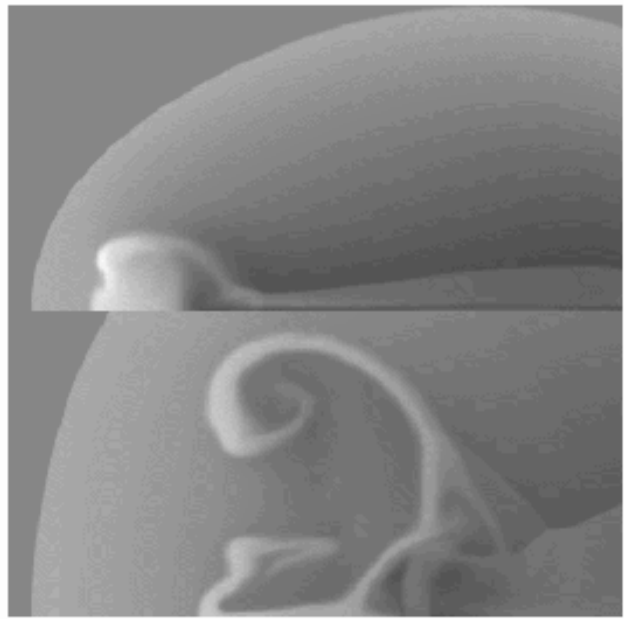
\item[Fig.~1] Grayscale images of the log density for models
T2 (a), T5 (b), A2 (c),
and A5 (d) at two times ($t/t_{bc} = 2,~6$) (top and bottom, respectively).
This shows bullet evolution for strong
and weak magnetic fields ($\beta_0 = 4,~256$)  and for both transverse
and aligned magnetic field geometries. Values increase from dark tones to
high tones.

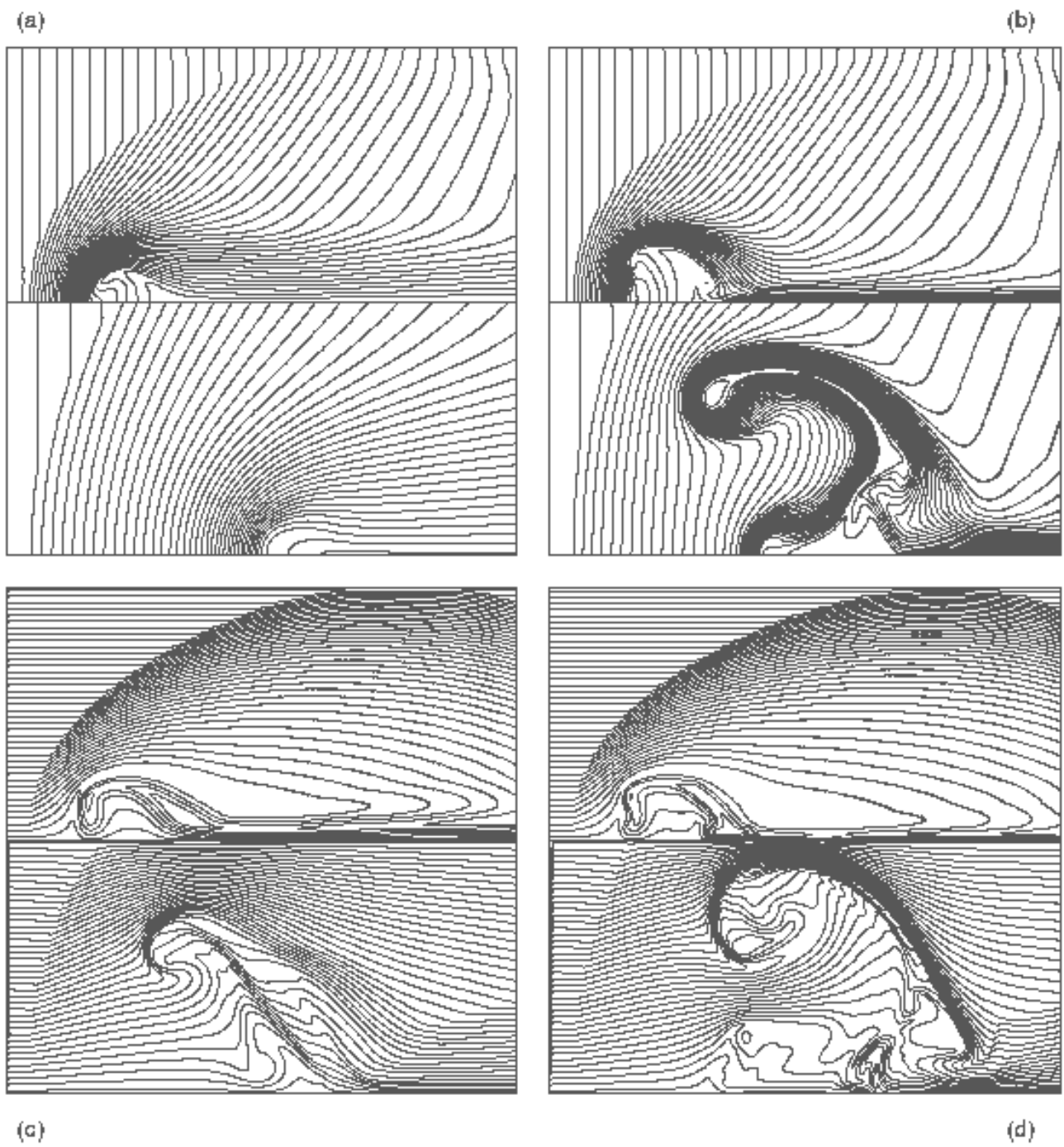
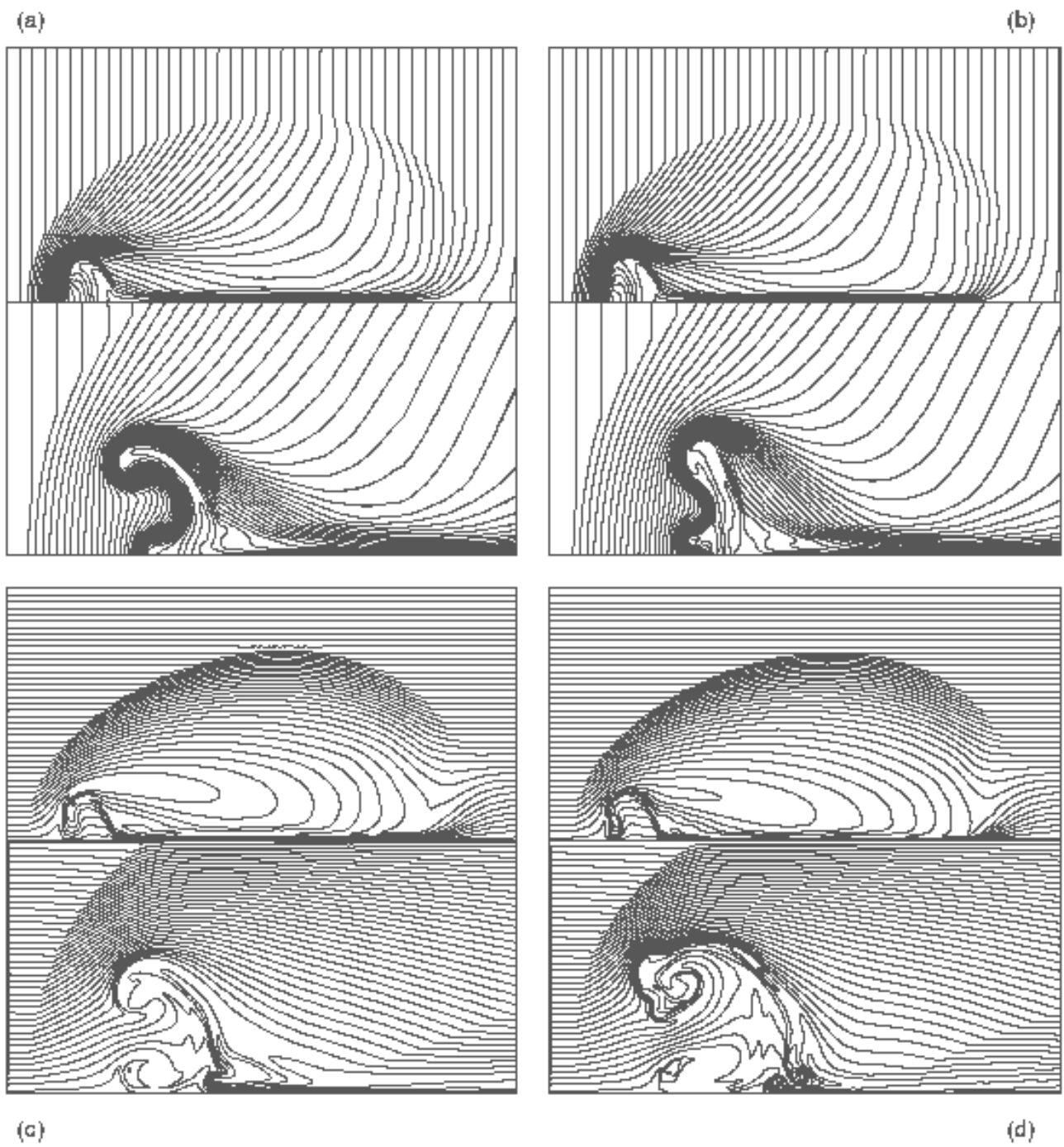
\item[Fig.~2] Same as Fig.~1, except showing the magnetic field
lines (contours of the magnetic flux function).
About 40 equal-interval contours of flux function are shown in each frame.

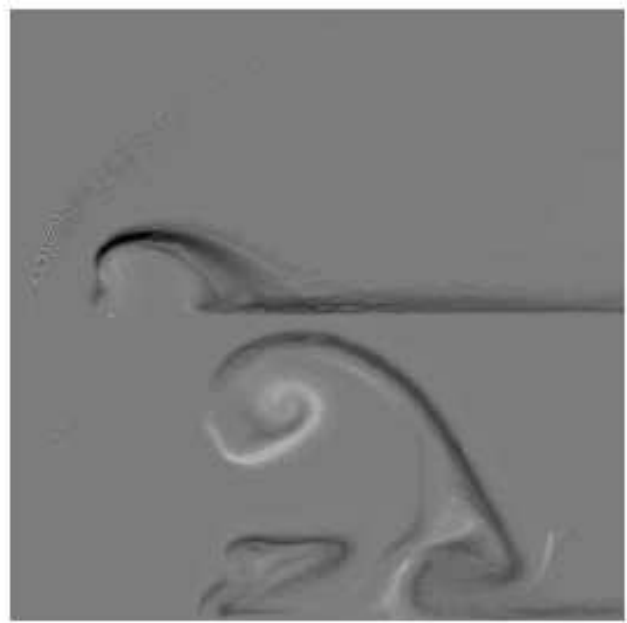
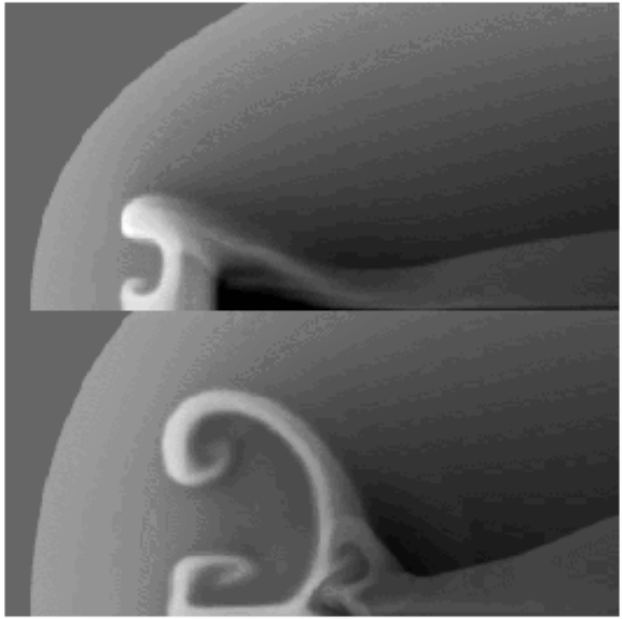
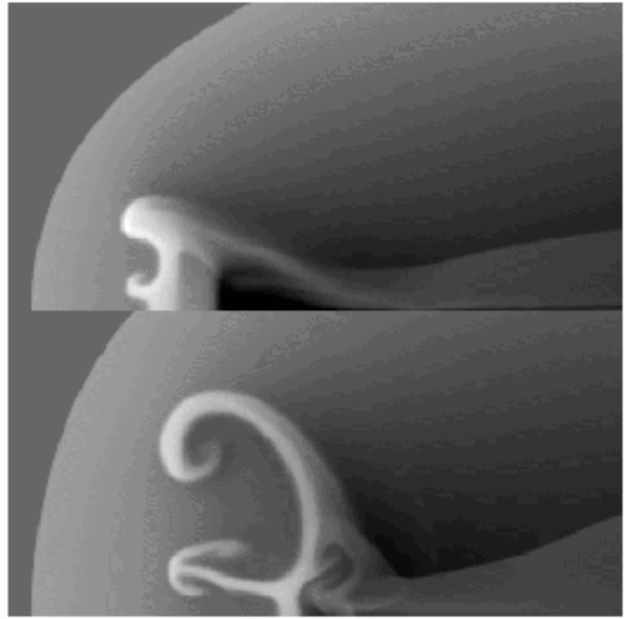
\item[Fig.~3] Same as Fig.~1, except showing the grayscale
images of  the vorticity. Dark 
tones are negative vorticity while high tones are positive.

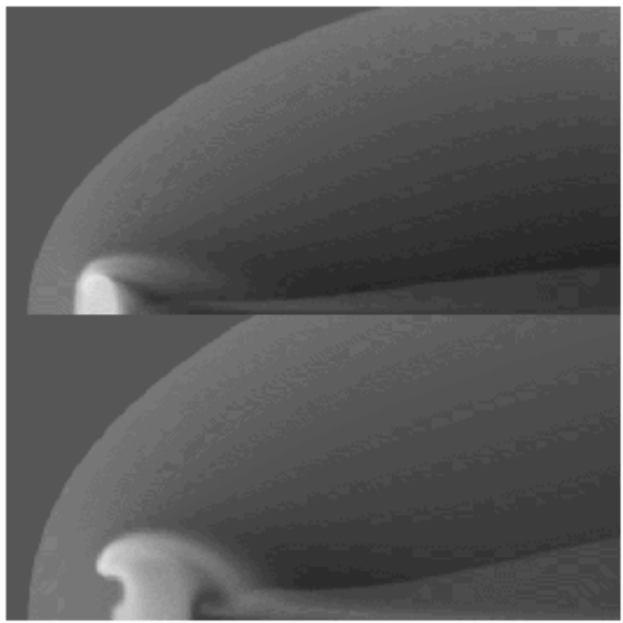
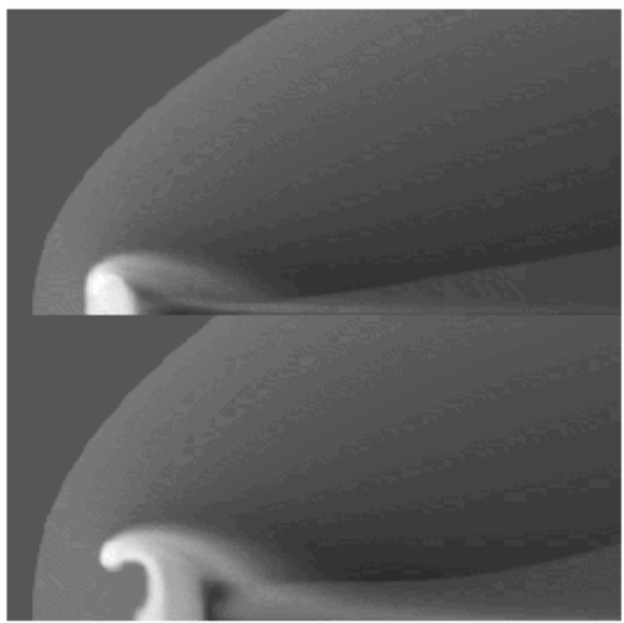
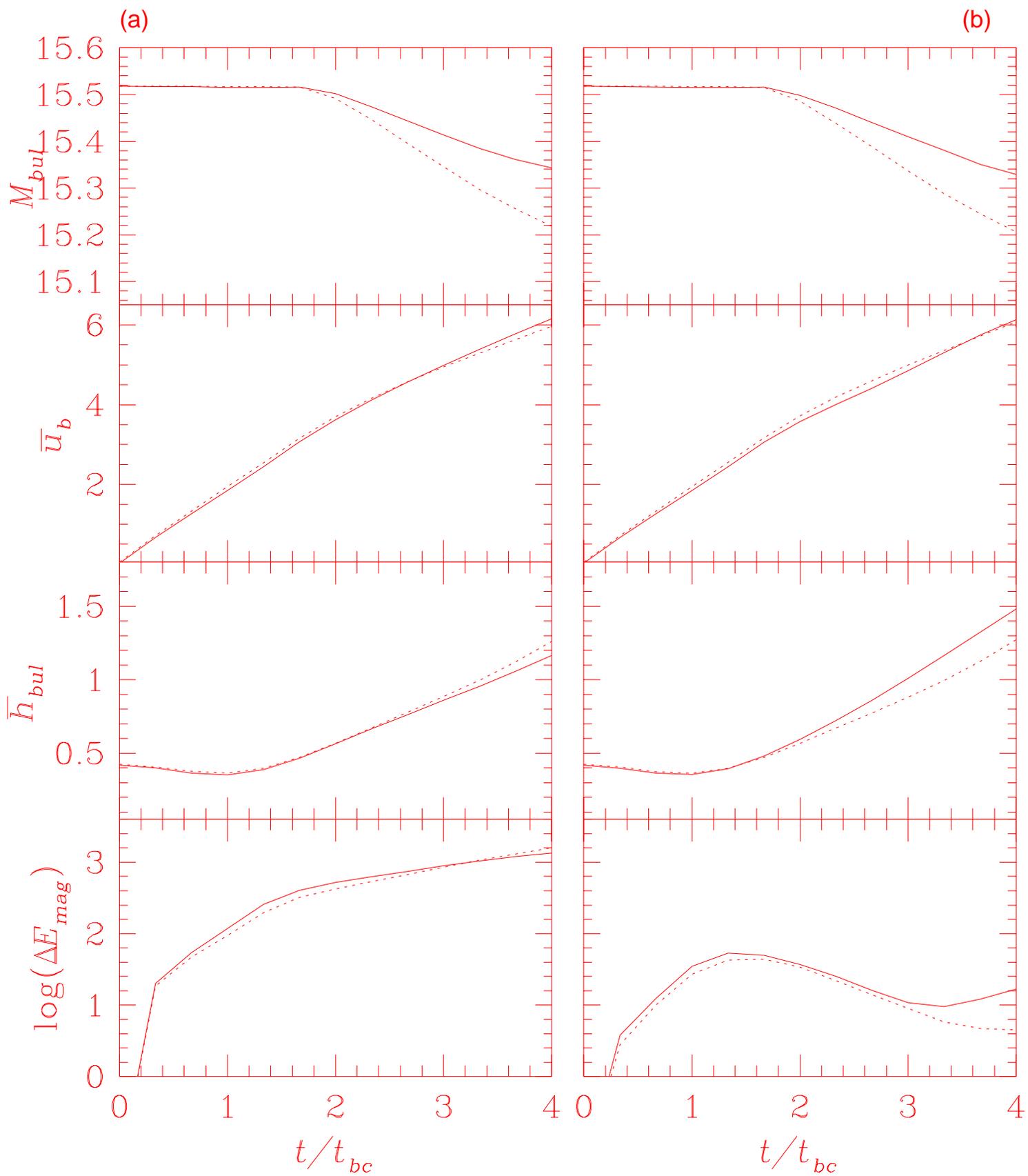
\item[Fig.~4] Grayscale images of the log density for gasdynamical
bullets ($\beta_0 = \infty$).
The three runs shown have $\chi = 10$, (a); $\chi = 40$, (b); and
$\chi = 100$, (c). Otherwise, the simulations are identical to MHD $R_{50}$
simulations. Each model is shown at $t/t_{bc} = 1$
(top) and $t/t_{bc} = 2$ (bottom).

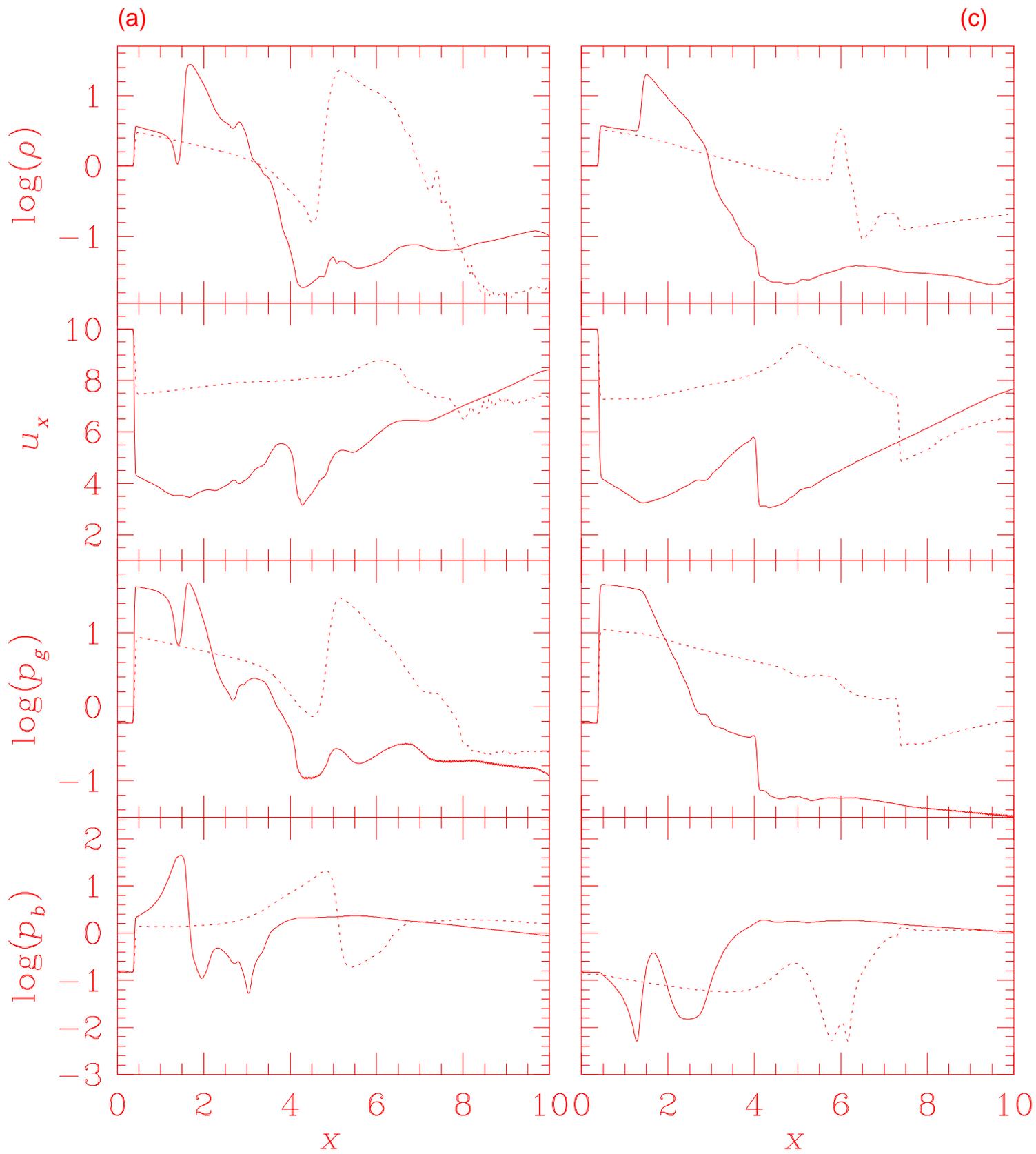
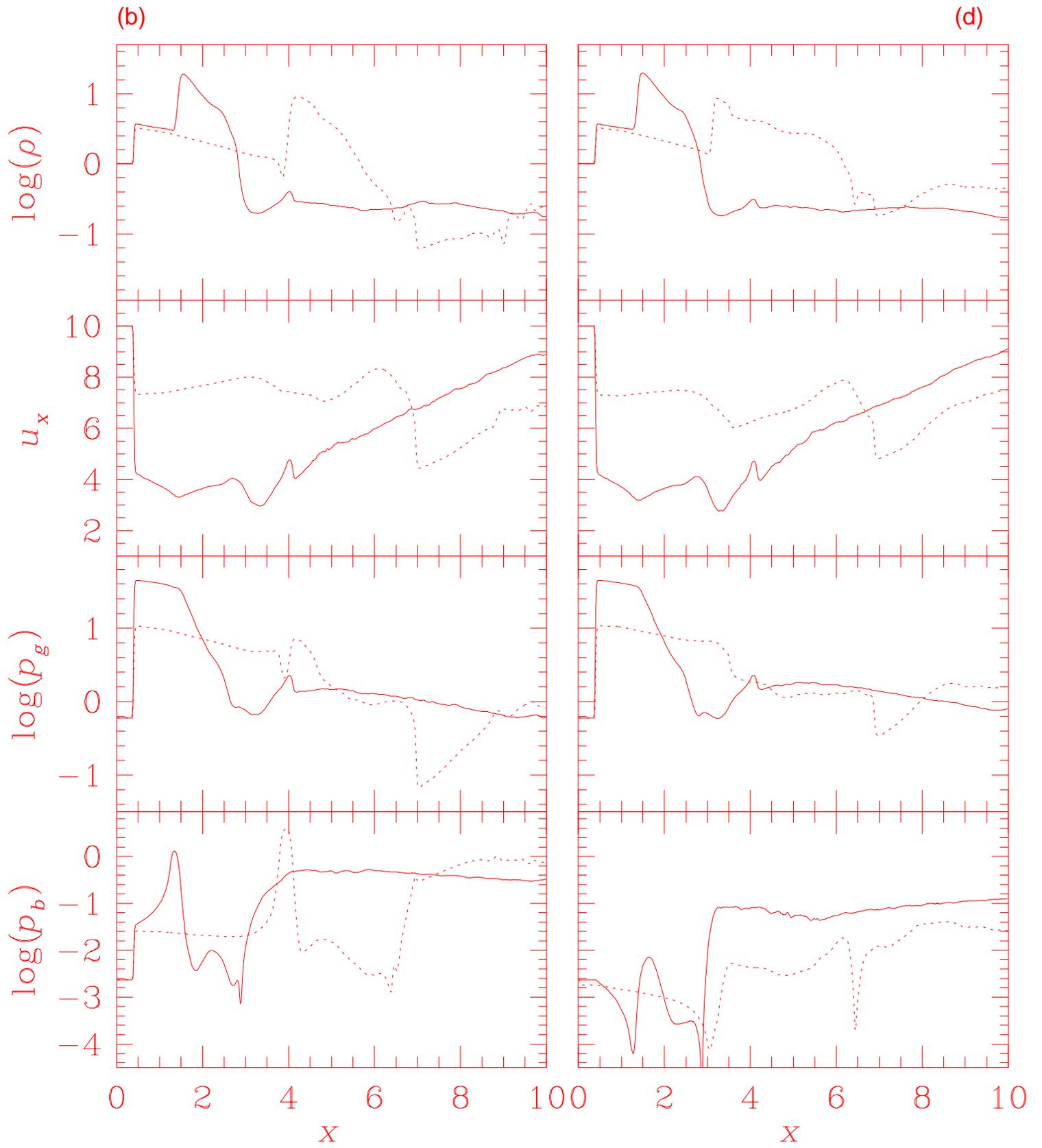
\item[Fig.~5] Cuts along $y=0.05$ (or the third cell from the bottom axis
with $y=0$) for models T2 (a), T5 (b), A2 (c), and A5 (d) at the same
time as in Fig.~1.
Solid lines correspond to $t/t_{bc} = 2$, while dotted lines represent
profiles at $t/t_{bc} = 6$. Shown top to bottom are: log gas density, $\rho$,
velocity as measured in the initial rest frame of the bullet, $u_x$,
log gas pressure, $p_g$, and log magnetic pressure, $p_b$.
nd log magnetic energy enhancement, $\Delta E_{mag}$ (Eq.~\ref{emag}).

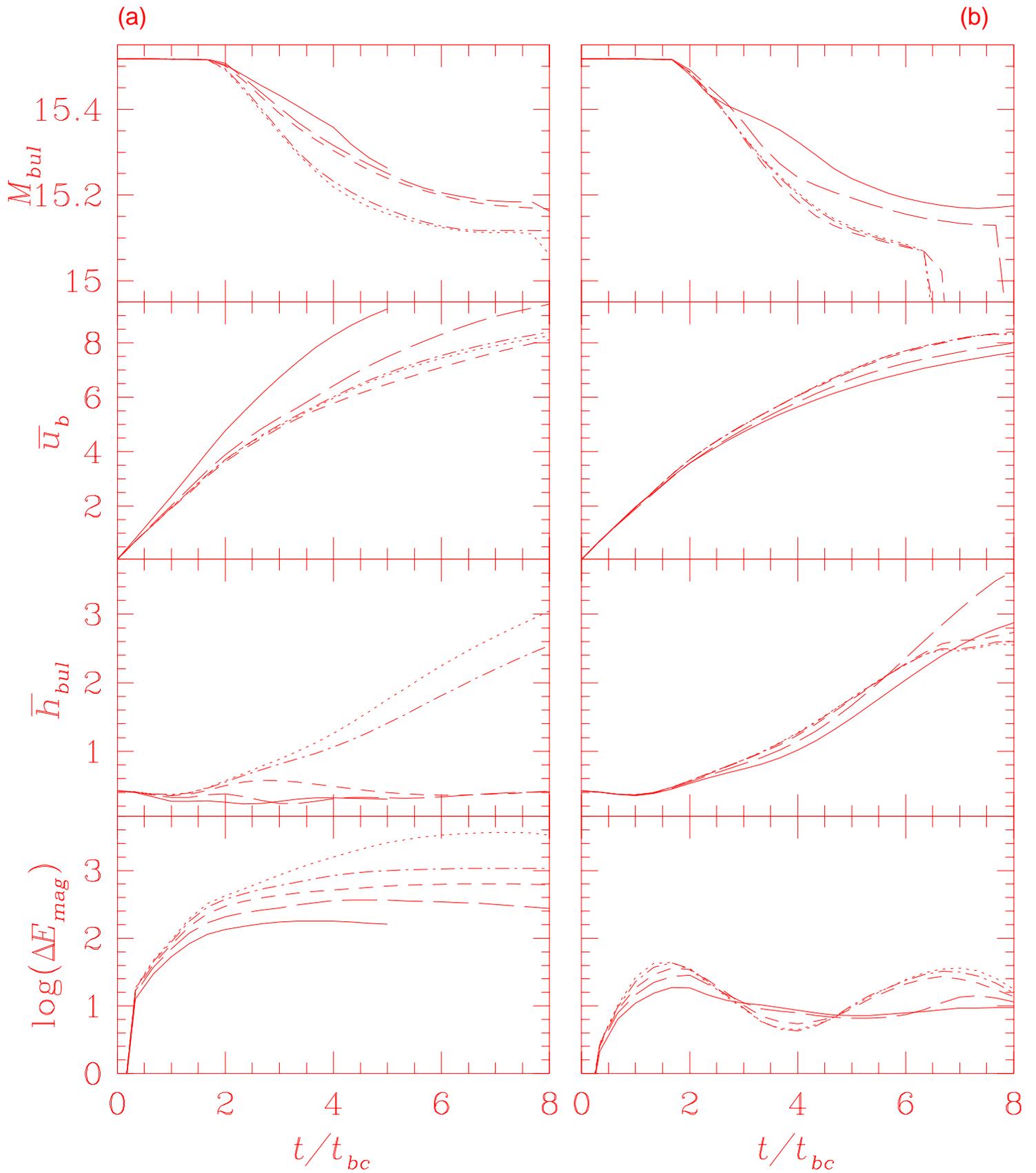
\item[Fig.~6] Evolution of various integrated quantities. Panel (a) presents
values for models with transverse magnetic fields (T1--T5), while panel (b)
for models with aligned fields (A1--A5). The line types
distinguish models of different initial field strength: solid ($\beta_0 = 1$),
long dash ($\beta_0 = 4$), short dash ($\beta_0 = 16$),
dot-dash ($\beta_0 = 64$),
and dot ($\beta_0 = 256$), with $\beta_0 = p_g/p_b$. Shown top to bottom
are: total bullet mass inside the grid, $M_{bul}$ (Eq.~\ref{mbull}),
mass-weighted bullet speed as measured in the initial rest frame
of the bullet, $\bar u_b$ (Eq.~\ref{uxav}),
mean height of the bullet material, $\bar h_{bul}$ (Eq.~\ref{yav}),
and log magnetic energy enhancement, $\Delta E_{mag}$ (Eq.~\ref{emag}).

\item[Fig.~a1] Grayscale images of the log density comparing
$R_{50}$ with $R_{100}$ runs for both magnetic field geometries when
$\beta_0 = 256$. Models T5 (a),
T6 (b), A5 (c) and A6 (d) are shown at $t/t_{bc} = {4}/{3}$ and $4$.

\item[Fig.~a2] Magnetic field lines for the same situations as Fig.~a1.
About 40 lines are shown in each frame.

\item[Fig.~a3] Grayscale images of the log density showing
sensitivity of bullet evolution to the initial details along
the symmetry axis $(y=0)$.  Panel (a) shows model A5 and panel (b)
an identical simulation with the initial bullet center shifted along
the $x$-axis ${1}/{2}$ zone. Times
shown are $t/t_{bc} = 3$ (top) and $t/t_{bc} = 5$ (bottom).

\item[Fig.~a4] Same quantities as Fig.~6, except now comparing
models with different resolution, T5 (dotted lines) and
T6 (solid lines) in (a) and A5 (dotted lines) and
A6 (solid lines) in (b).

\end{description}

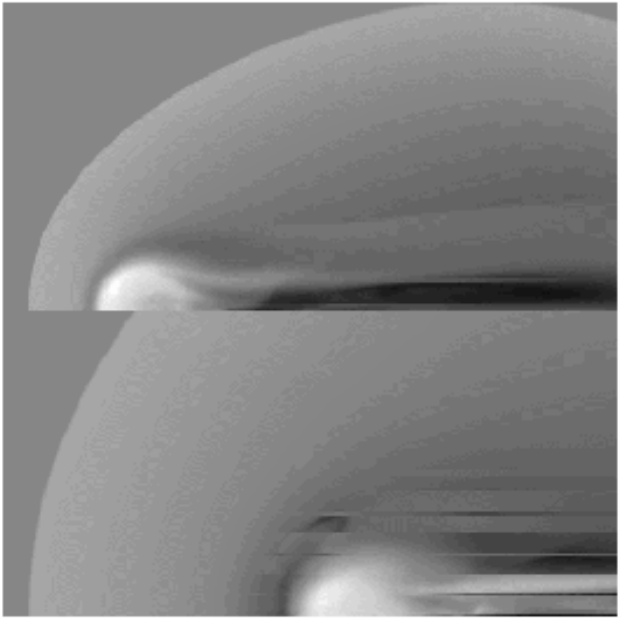
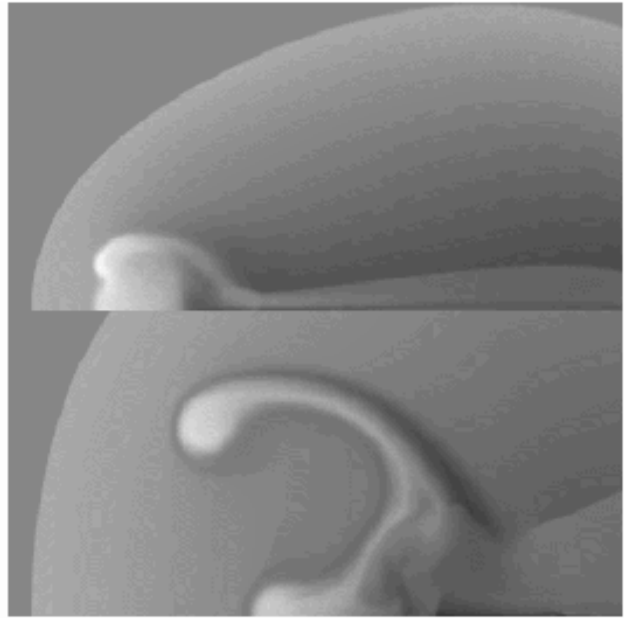
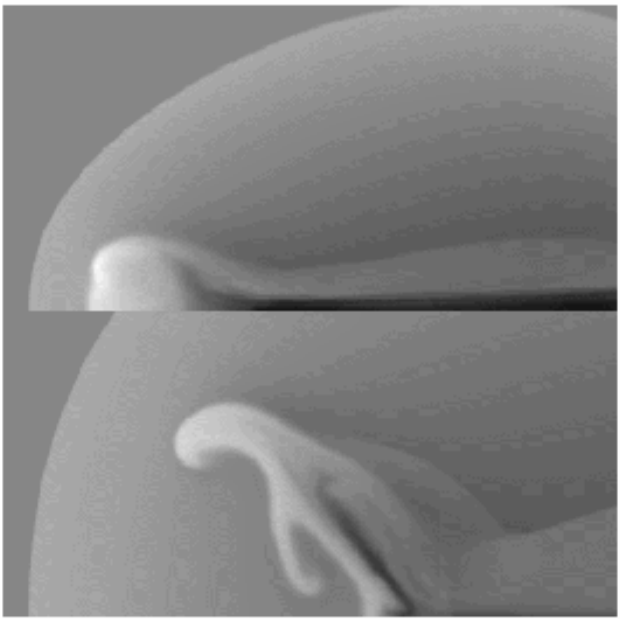
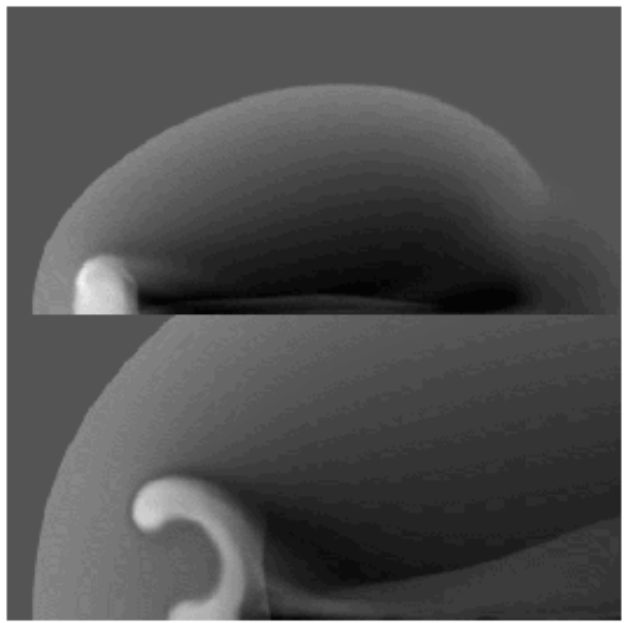
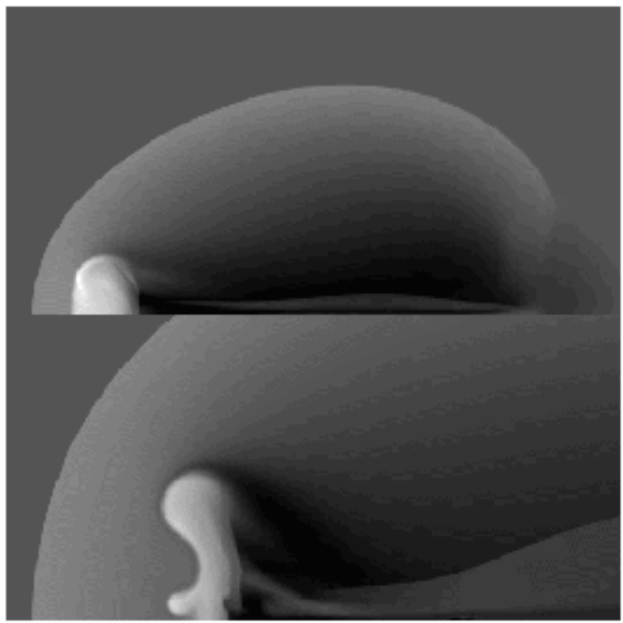
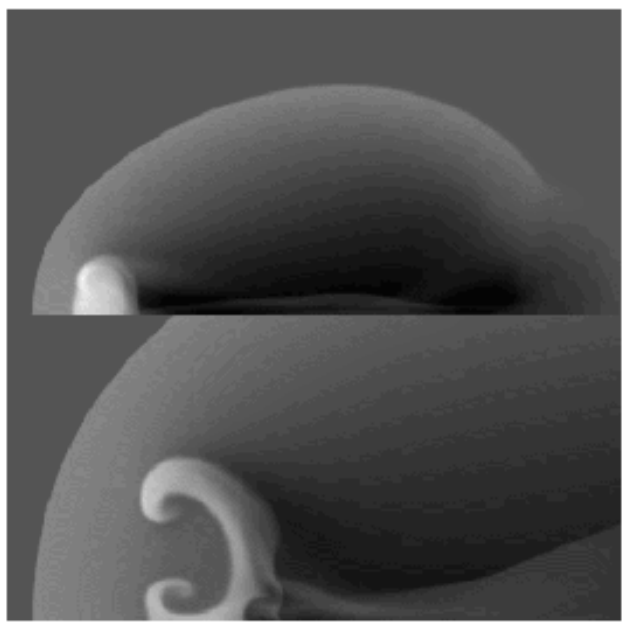
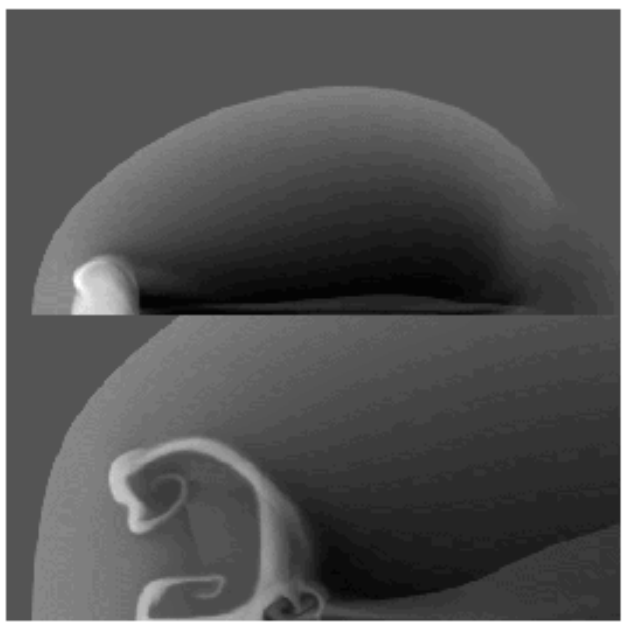
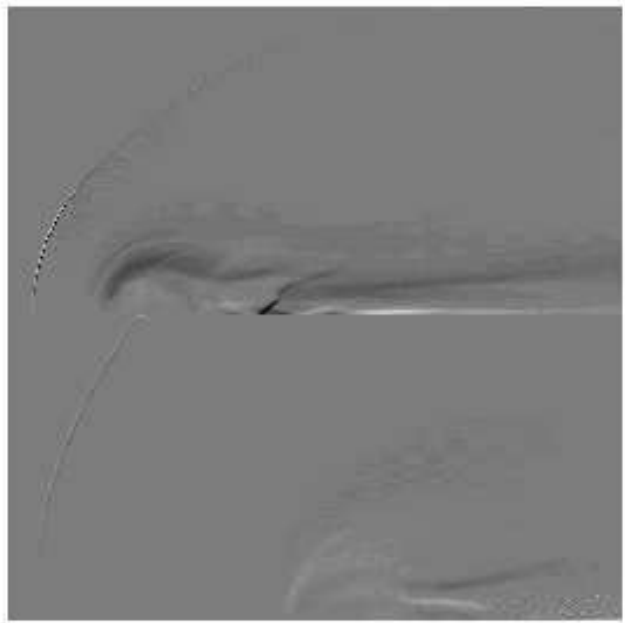
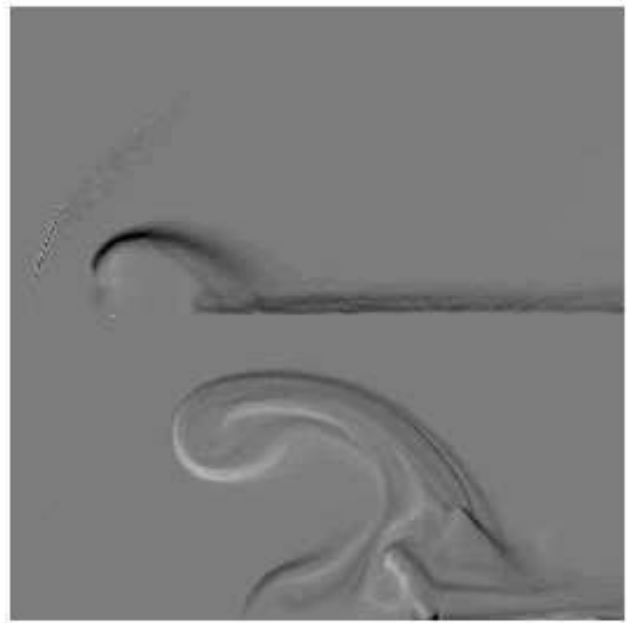
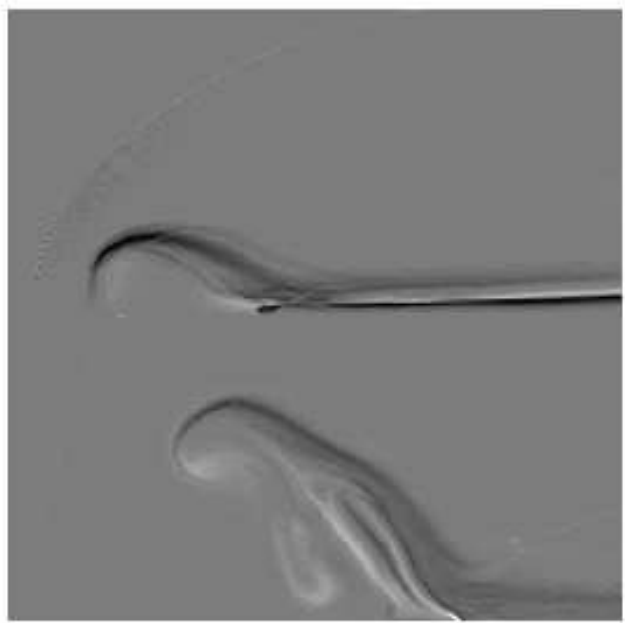
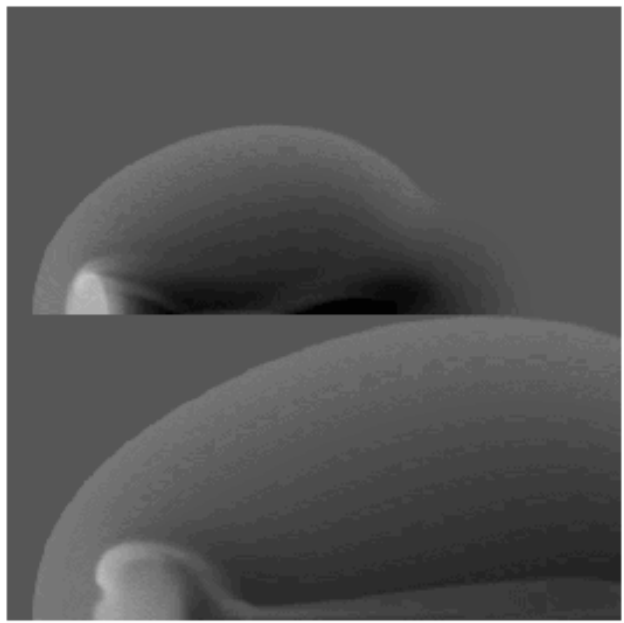
\end{document}